\documentclass[preprint,usenatbib]{aastex631}
\usepackage[utf8]{inputenc}
\usepackage{amsmath}
\usepackage{mathtools}
\usepackage{amsfonts}
\usepackage{amssymb}
\usepackage{makeidx}
\usepackage{indentfirst}
\usepackage{graphicx}
\usepackage{color}
\usepackage{lmodern}
\usepackage{url}
\usepackage{soul}
\usepackage{mathrsfs}
\usepackage{longtable}
\usepackage{multirow}
\usepackage [english]{babel}
\usepackage [autostyle, english = american]{csquotes}
\MakeOuterQuote{"}
\usepackage{chemformula}
\usepackage{textcomp}
\usepackage{aas_macros}
\usepackage{natbib}
\usepackage{tabularx}
\usepackage{comment}
\usepackage{soul}
\usepackage{xcolor}
\definecolor{voilet}{RGB}{127,0,255}
\definecolor{vikas1}{RGB}{255,198,0}
\definecolor{vikas2}{RGB}{0, 255, 0}
\definecolor{vikas3}{RGB}{0, 0, 255}
\definecolor{vikas}{RGB}{255,0,127}
\definecolor{KA}{rgb}{0.36, 0.73, 0.58}

\shorttitle{Vertical mixing in exoplanet atmospheres}
\shortauthors{Soni \& Acharyya}

\begin{document}
\title{Signature of Vertical Mixing in Hydrogen-dominated Exoplanet Atmospheres}

\author[0000-0001-9273-9694]{Vikas Soni}
\affiliation{Planetary Sciences Division, Physical Research Laboratory, Ahmedabad, 380009, India}

\author[0000-0002-0603-8777]{Kinsuk Acharyya}
\affiliation{Planetary Sciences Division, Physical Research Laboratory, Ahmedabad, 380009, India}


\begin{abstract}
Vertical mixing is a crucial disequilibrium process in exoplanet atmospheres, significantly impacting chemical abundance and observed spectra. While current state-of-the-art observations have detected its signatures, the effect of vertical mixing on atmospheric spectra varies widely based on planetary parameters. In this study, we explore the influence of disequilibrium chemistry across a parameter space that includes eddy diffusion, surface gravity, internal and equilibrium temperature, and metallicity. We also assess the effectiveness of retrieval models in constraining the eddy diffusion coefficient. By running numerous 1D chemical kinetics models, we investigate the impact of vertical mixing on the transmission spectrum. We also built a custom fast-forward disequilibrium model, which includes vertical mixing using the quenching approximation and calculates the model abundance orders of magnitude faster than the chemical kinetics model.  We coupled this forward model with an open source atmospheric retrieval code and used it on the JWST simulated output data of our chemical kinetics model and retrieved eddy diffusion coefficient, internal temperature and atmospheric metallicity. We find that there is a narrow region in the parameters space in which vertical mixing has a large effect on the atmospheric transmission spectrum. In this region of the parameter space, the retrieval model can put high constraints on the transport strength and provide optimal exoplanets to study vertical mixing. Also, the \ch{NH3} abundance can be used to constrain the internal temperature for equilibrium temperature $T_{\text{equi}} > 1400$ K.
\end{abstract}

\section{Introduction}
Deciphering the atmospheric composition of exoplanets is crucial to understanding the physical and chemical processes occurring in the exoplanet atmospheres. The disequilibrium processes, such as vertical mixing and photochemistry, play a significant role in governing atmospheric properties. With the advent of state-of-the-art observational facilities like JWST, the characterization of the atmosphere of exoplanets has entered a phase where the observed spectral signature can probe the disequilibrium chemistry in the atmosphere. Concrete evidence of both photochemistry and vertical mixing has recently been found. Detection of \ch{SO2} using JWST provided evidence of photochemistry for the first time \citep{Alderson2023, Rustamkulov2023}.  Likewise, \cite{Baxter2021} found evidence of the vertical transport of \ch{CH4} while analyzing the transit curve of 49 gas giants. Recently, \cite{Sing2024Natur} also found the signature of vertical mixing in WASP-107 b.

The retrieval models are crucial to determining the atmospheric properties from the observed spectra. When radiation passes through the exoplanet atmosphere and reaches the observer, signatures of various physical and chemical processes such as chemical composition, temperature structure, atmospheric circulation, and clouds/hazes are imprinted on it. For a given spectrum, extracting various components with their uncertainty is the goal of the retrieval model. Typically, a large number of forward model runs is required to perform retrieval, and this number depends upon the spectral resolution and the parameter of interest, which is included in the retrieval \citep{Madhusudhan2011, Madhusudhan2019}. The purpose of these retrieval models is to extract the relevant information and put statistical constraints on the parameters (mixing ratio, elemental abundance, thermal profile, disequilibrium processes) of interest from the observed spectrum.

To date, a large number of retrieval models have been developed and applied to decode the secrets of exoplanet atmospheres (\citealp{Madhusudhan2009, Madhusudhan2010,  Madhusudhan2011a, Madhusudhan2014, Benneke2013, Line2013, Line2014, Cubillos2015, Waldmann2015b, Waldmann2015a, MacDonald2017b, Wakeford2017, Gandhi2018, Welbanks2021, Kawashima2021, Nixxon2022, Chubb2022}, and references therein). The constant abundance profile (free chemistry retrieval) or the thermochemical equilibrium abundance (chemically consistent retrieval) are widely incorporated in the retrieval models. However, incorporating disequilibrium processes has been very challenging, though evidence suggests that they are indispensable components in understanding atmospheric compositions \citep{Prinn1977, Moses2011,Moses2013, Moses2016, Venot2014, Venot2020, Morley2017, Tsai2017, Tsai2021, Fortney2020, Molliere2020, Baxter2021}. Recently, some attempts have been made to include disequilibrium chemistry in the retrieval. \cite{Morley2017} used a free parameter for the quench level and a single quench level for all the species. \cite{Molliere2020} calculated the \ch{CH4} and CO quench levels in their retrieval model by making use of the analytical timescales from \cite{Zahnle2014}. \cite{Ahmed2024} implemented a full chemical kinetic scheme into an atmospheric retrieval framework using the distillation method. \cite{Kawashima2021} used the chemical relaxation method to include vertical mixing in the retrieval model and found an indication of disequilibrium chemistry for HD 209458 b. 

The effect of vertical mixing on the chemical abundance of the atmosphere differs from planet to planet and mainly depends upon the planetary thermal profile. For hot exoplanets, the chemical timescale is faster than the dynamical timescale, resulting in an atmospheric composition that is governed by the thermochemical equilibrium abundance. For very cold exoplanets, the thermal profile can entirely lie in the \ch{CH4} or \ch{NH3} dominant region, and the entire atmosphere follows the constant mixing ratio profile. In this case,  vertical mixing does not affect the abundance of \ch{CH4} and \ch{NH3}.

 Theoretical and observational studies indicate a narrow range of atmospheric parameters of exoplanets, also sometimes referred to as a 'sweet spot' \citep{Zamyatina2023}, for which the signature of vertical mixing in the observed spectrum is maximum. \cite{Kawashima2021} also found the signature of disequilibrium chemistry in HD 209458 b and WASP-39 b while performing the spectral retrieval of the transmission spectra of 16 exoplanets. 

The characteristics of this sweet spot for the observability of vertical mixing are yet to be fully understood. How atmospheric retrieval models, which include vertical mixing, will constrain the strength of vertical mixing in this parameter space has also not been studied previously. In this work, our goal is to examine how efficiently we can constrain vertical mixing using the quenching approximation in an atmospheric retrieval model. The methodology for including the quenching approximation in the retrieval model is discussed in Section \ref{sec:quench}. Section \ref{sec:sign_v_mix} contains the results of the signature of vertical mixing in the transit spectrum. The retrieval results are presented in Section \ref{sec:rslt}. The discussions are given in Section \ref{sec:discuss}, and the concluding remarks are made in Section \ref{sec:concl}.

\section{Quenching approximation in the atmospheric retrieval model}\label{sec:quench}
The quench level is defined at a pressure level where the chemical conversion timescale is equal to the vertical mixing timescale of the atmosphere. Below the quench level, the chemical timescale dominates over the transport timescale, and the atmospheric composition remains at chemical equilibrium. Above the quench level, the abundance of a species freezes at the equilibrium abundance at the quench level  \citep{Prinn1977,Lodders2002,Zahnle2014,Tsai2018}. The quenching approximation as described by \cite{Smith1998} can give an accurate result within 10\% of the kinetics/transport model \citep{Moses2011}. This approximation is routinely used in various studies to constrain the disequilibrium abundance of exoplanets \citep{Line2010, Madhusudhan2011, Moses2011, Visscher2012, Zahnle2014, Tsai2017, Tsai2018, Fortney2020, Soni2023a, Soni2023b, Zamyatina2024}. We have used the quenching approximation 
to include vertical mixing in the retrieval model to determine the abundance of assorted H-C-N-O bearing species. We use the abundance to constrain the eddy diffusion coefficient $K_{zz}$, the internal temperature $T_{\text {int}}$ and the metallicity [M/H] of a planet.

\subsection{Vertical Mixing Timescale} \label{mix}
The vertical mixing timescale $\tau_{\text{mix}}$ can be computed using the mixing length theory, and is given by the following equation:
\begin{equation} 
\tau_{\text{mix}} = L^2 / K_{zz},
\end{equation}
where $L$ is the mixing length scale of the atmosphere \citep{Visscher2011, Heng2017}. The mixing length scale cannot be computed from the first principle, and a simple approximation is to take the pressure scale height as the mixing length. We use the method described by \cite{Smith1998} in which it is found that the mixing length can be $L \approx 0.1-1 \times \text{pressure scale height}$, which leads to $\tau_{\text{mix}} = (\eta H)^2 / K_{zz}$, where $\eta \in[0.1,1]$ and the exact value of $\eta$ depends upon the rate of change of chemical timescale with height. The pressure scale height $H = \frac{k_B  T}{\mu g}$, where $k_B$ is the Boltzmann constant and $T$, $g$ and $\mu$ are the temperature, surface gravity, and mean molecular mass of the atmosphere, respectively.

\subsection{Chemical Timescale}\label{Chem}
In chemical equilibrium, the abundance of the chemical species does not change with time and the chemical reactions take place in a way that the production and loss rates of any species are balanced, such that
\begin{equation}
\frac{dn_{\text{EQ}}}{dt} = P_{EQ} - n_{\text{EQ}} L_{EQ}= 0.
\end{equation} 
Here, $n_{\text{EQ}}$, $P_{\text{EQ}}$ and $n_{\text{EQ}} L_{\text{EQ}}$ are the number density, production rate and loss rate, respectively, and $L_{\text{EQ}}$ is independent of $n_{\text{EQ}}$. When the physical conditions such as temperature and pressure of the system change, or the species are transported into other regions of the atmosphere, the chemical abundance deviates from chemical equilibrium, and the species are converted among themselves to restore chemical equilibrium. The conversion of species takes place through several chains of chemical reactions, which are called conversion schemes. The timescales of reactions in a conversion scheme can vary significantly and the conversion timescale is computed from the slowest reaction in the fastest conversion scheme, known as the rate-limiting step (RLS)  \citep{Visscher2011, Madhusudhan2011, Zahnle2014, Tsai2017}. The timescale of the conversion of species $a$ into $b$ is 
\begin{equation}
\tau_{a\rightarrow b} = \frac{[a]}{\text{Rate of RLS}_{a\rightarrow b}}.
\end{equation}
Here, [a] is the abundance of species $a$, and $\text{RLS}_{a\rightarrow b}$ is the rate-limiting step in the conversion of $a$ into $b$. In a chemical network, a particular species is involved in several reactions and there are many conversion pathways between two species. The number of these pathways increases exponentially with an increase in the number of reactions in the network. However, only a few conversion schemes are important in a chemical network, as most of them are significantly slower than the fastest conversion scheme. 

The timescale of the reactions is a function of several parameters, especially temperature and pressure. As these parameters change, the rate-limiting step (RLS) also changes, i.e., a reaction that is the RLS for a given set of parameters may not be the RLS as these parameters get altered. Thus, the rate-limiting reaction can change in the parameter space. We developed a chemical network analysis tool which takes a chemical network, finds all the possible conversion schemes between the species and then finds the RLS.

\subsection{Methodology to Apply the Quenching Approximation in the Retrieval Model}
We developed a Python-based disequilibrium model function that calculates the chemical abundances of the atmosphere using the quenched curve data of \ch{CH4}, \ch{CO}, \ch{CO2}, \ch{H2O}, \ch{NH3}, \ch{N2}, and \ch{HCN} from our previous work \cite{Soni2023a} and \cite{Soni2023b}. The quenched curve finds the initial quench pressure on an input thermal profile for the respective species. These initial quench levels are then further fine-tuned by the Smith method \citep{Smith1998} to find the quenched abundance of the species. 

For \ch{CH4}, CO, \ch{H2O}, \ch{NH3}, and \ch{N2}, the abundances in our disequilibrium model function are taken as the thermochemical equilibrium abundances below their quench levels and the frozen mixing ratios above their quench levels. However, \ch{CO2} and \ch{HCN} remain in pseudo-equilibrium with CO, \ch{H2O}, and \ch{NH3}. For the pressure higher than their quench level pressure, we have used the following equation for the abundance of \ch{CO2} \citep{Moses2011,Tsai2018,Soni2023a}:
\begin{equation}
[\ch{CO2}] = k \frac{[\ch{CO}_q][\ch{H2O}_{q}]}{[\ch{H}_{2,eq}]}
\end{equation}
and the HCN abundance is calculated as follows \citep{Soni2023b}:
\begin{equation}
[\ch{HCN}] = k \frac{[\ch{NH}_{3,q}][\ch{CO}_q][\ch{H}_{eq}]}{[\ch{OH}_{eq}][\ch{H}_{2,eq]}},
\end{equation}
where $[\ch{CO}_q]$, $[\ch{H2O}_{q}]$, and $[\ch{NH}_{3,q}]$ are the quenched abundance of \ch{CO}, \ch{H2O}, and \ch{NH3}, respectively. $[\ch{H}_{eq}]$, $[\ch{H}_{2,eq}]$, and $[\ch{OH}_{eq}]$ are the equilibrium abundances of \ch{H}, \ch{H2} and \ch{OH}, respectively. 
Above the quench level of \ch{CO2} and \ch{HCN}, we use their quenched mixing ratios. 

\begin{figure}[t!]
	\centering
	\includegraphics[width=0.8\linewidth]{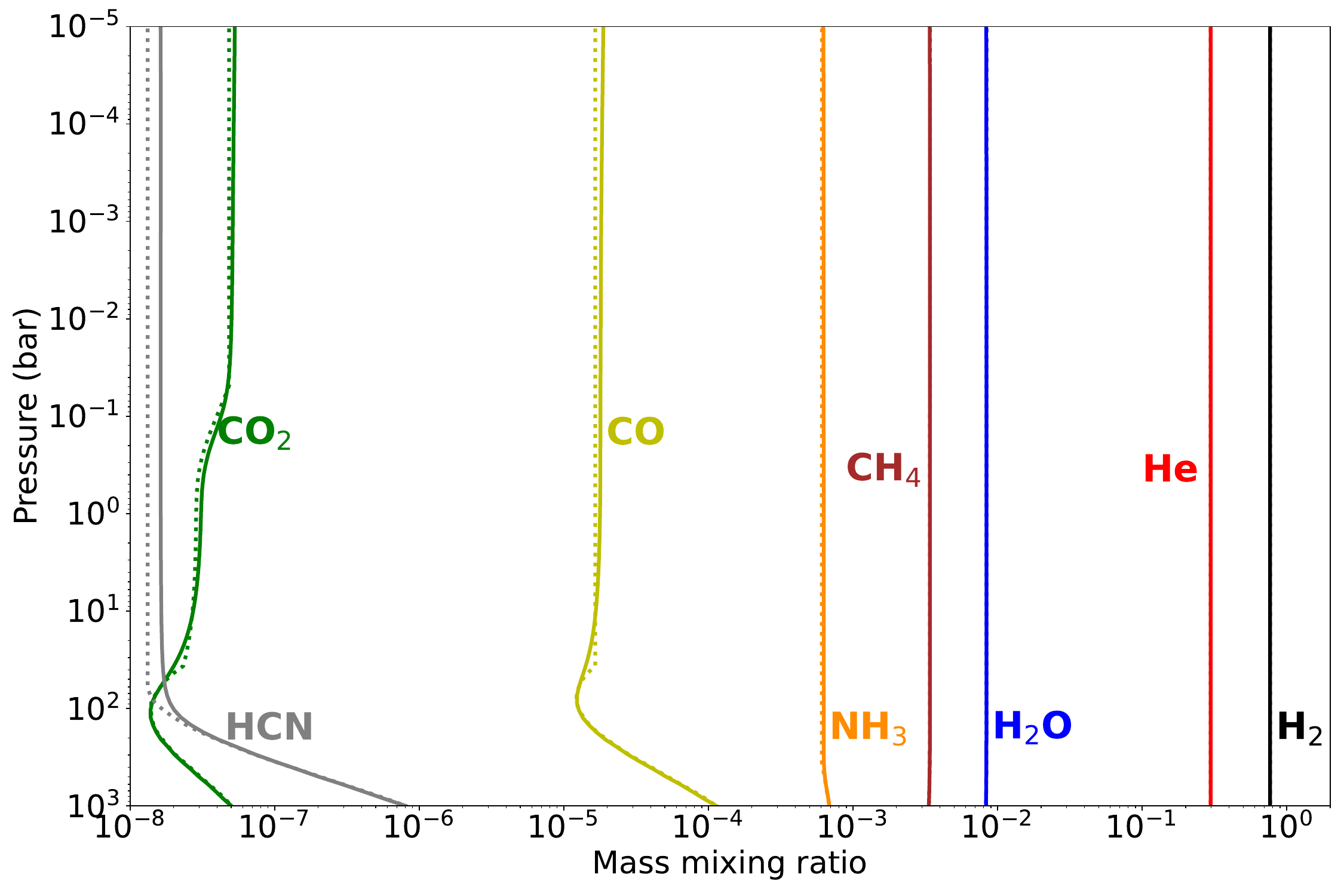}
	\caption{ The mass mixing ratios for a test exoplanet (model B in the Appendix: $\text{log}_{10}(g/\text{cm} \text{ s}^{-2})$ = 3, $\text{log}_{10}(K_{zz}/\text{cm}^2 \text{ s}^{-1})$ = 9, [M/H] = 0, internal temperature = 200 K and equilibrium temperature = 1000 K ). The solid lines are the output of the 1D chemical kinetics model and the dashed lines are the output of the disequilibrium model function.}\label{Fig:CO_compare}
\end{figure}

The disequilibrium model function takes a thermal profile, eddy diffusion coefficient, surface gravity,  metallicity, and a list of the species whose abundance needs to be calculated as the input parameters. The model returns a Python dictionary with the abundance in the mass fraction of the species present in the input species list. The typical computation time is about 20 $\mu$s, several orders of magnitude faster than the 1D chemical transport model.  
Figure \ref{Fig:CO_compare} shows the agreement between the output of the 1D chemical kinetics model and the output of the disequilibrium model function, which uses the quenching approximation to calculate the abundance. We then use the disequilibrium model function in an open-source retrieval model (petitRADTRANS; \citealt{Molliere2019}) as a model generation function; in short, we call this the quench level retrieval (QLR). 
\begin{figure}[t!]
	\centering
	\includegraphics[width=1\linewidth]{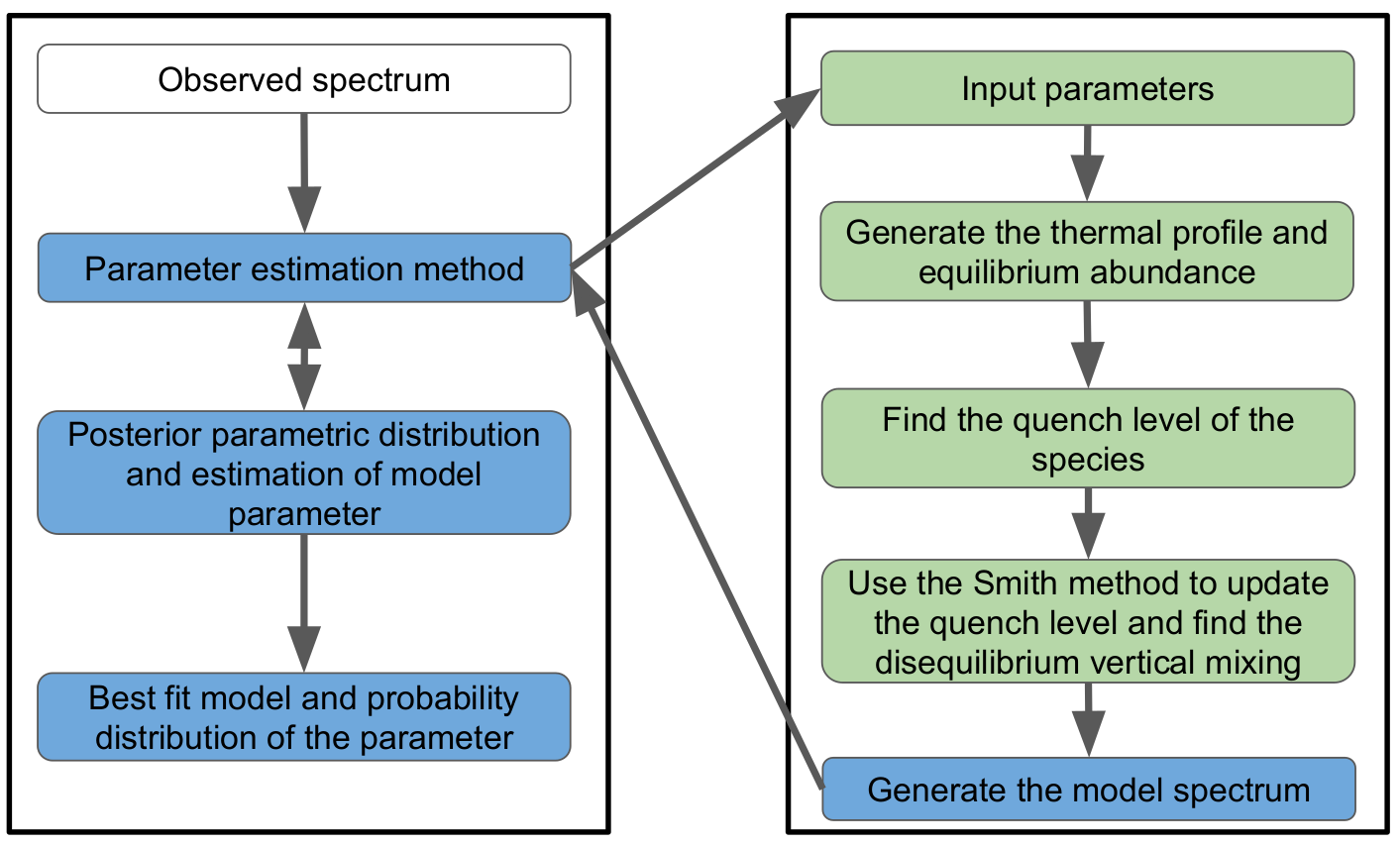}
	\caption{Block diagram of the quench level retrieval (QLR). The model calculates the output spectrum for an input planet spectrum and its parameters. The parameter estimation method (PyMultiNest) is used to find the parameter distribution and calculate the best-fit parameters for the input spectrum. The blue boxes are the steps that are done by petitRADTRANS, and the green boxes are the steps carried out by our disequilibrium model function.}\label{Fig:block_diagram}
\end{figure}

The petitRADTRANS retrieval module is an easy-to-use and versatile Python package for primary and secondary transit retrieval calculations. petitRADTRANS includes two modes of spectra calculation: the low-resolution mode (R = $\frac{\lambda}{\delta \lambda}$ = 1000) uses the correlated-k method, and the high-resolution mode (R = 10$^6$) uses the line-by-line method. It includes the line opacity of \ch{H2O}, \ch{O3}, \ch{OH}, \ch{CO}, \ch{CO2}, \ch{CH4}, \ch{C2H2}, \ch{NH3}, HCN, \ch{H2S}, \ch{PH3}, \ch{TiO}, \ch{VO}, \ch{SiO}, \ch{FeH}, \ch{H2}, \ch{Na}, \ch{K} and also incorporates the pressure broadening.
For the low-resolution mode, the range of line opacity is $0.3-28$ $\mu$m, while it is 110 nm $-$ 250 $\mu$m for the high-resolution mode. These opacity values are available for a temperature range of $80 - 3000$ K and a pressure range of $10^{-6} - 10^3$ bar. Along with the line opacity, petitRADTRANS also includes Rayleigh scattering (for \ch{H2}, \ch{He}, \ch{H2O}, \ch{CO2}, \ch{CO}, \ch{O2}, \ch{CH4}, and \ch{N2}) and cloud opacity (for \ch{Al2O3}, \ch{H2O}, \ch{Fe}, \ch{KCl},  \ch{MgAl2O4}, \ch{MgSiO3}, \ch{Mg2SiO4}, and \ch{Na2S} clouds). petitRADTRANS is benchmarked with the ATMO \citep{Tremblin2015} and Exo-REM \citep{Baudino2015} codes. To sample the parameter space in the Bayesian statistics, petitRADTRANS uses the nested sampling algorithm from PyMultiNest. 

The block diagram of the QLR is shown in Figure \ref{Fig:block_diagram}. The QLR requires an input spectrum and a list of free parameters with their prior values. The parameters can also be fixed at a certain value. The QLR uses the initial retrieval run by a number of initializers and finds the best-fit model using PyMultiNest, which uses the nested sampling method to find the best-fit parameters in good agreement. We have discussed the benchmarking of the model in Appendix.

\section{Vertical mixing in the exoplanet atmosphere}\label{sec:sign_v_mix}

Vertical mixing causes the mixing of the chemical species from one pressure region to another. As a result, in the reference frame of the species, its local temperature and pressure change. In these new pressure and temperature conditions, the thermochemical equilibrium abundance of the species can be different. Thus, the abundance of the transported species starts to change towards its new thermochemical equilibrium abundance. In this conversion, the elements of the species are transferred to other stable species. The steady-state mixing ratio of the chemical species depends upon the strength of vertical mixing, the chemical conversion timescale and the thermal profile of the atmosphere. The shape and location of the thermal profile in the temperature-pressure space largely affect the extent of the effect of vertical mixing in the atmospheric composition of the atmosphere, as the thermal profile decides what will be the thermochemical equilibrium abundance in different parts of the atmosphere. The thermal profile also controls the chemical timescale and the abundance of species at the quench level. For sufficiently high-temperature exoplanets (effective temperature $ > 2000$ K), the effect of vertical mixing is negligible as the chemical timescale is fast compared to the dynamical timescale of the atmosphere, which favors equilibrium chemistry. In addition, the thermal profile entirely lies in the \ch{CO} and \ch{N2} dominant region, making \ch{CO} and \ch{N2} the major species of C and N throughout the atmosphere  \citep{Lodders2002,Moses2013}.  

\begin{figure}[b!]
\centering
\includegraphics[width=1\textwidth]{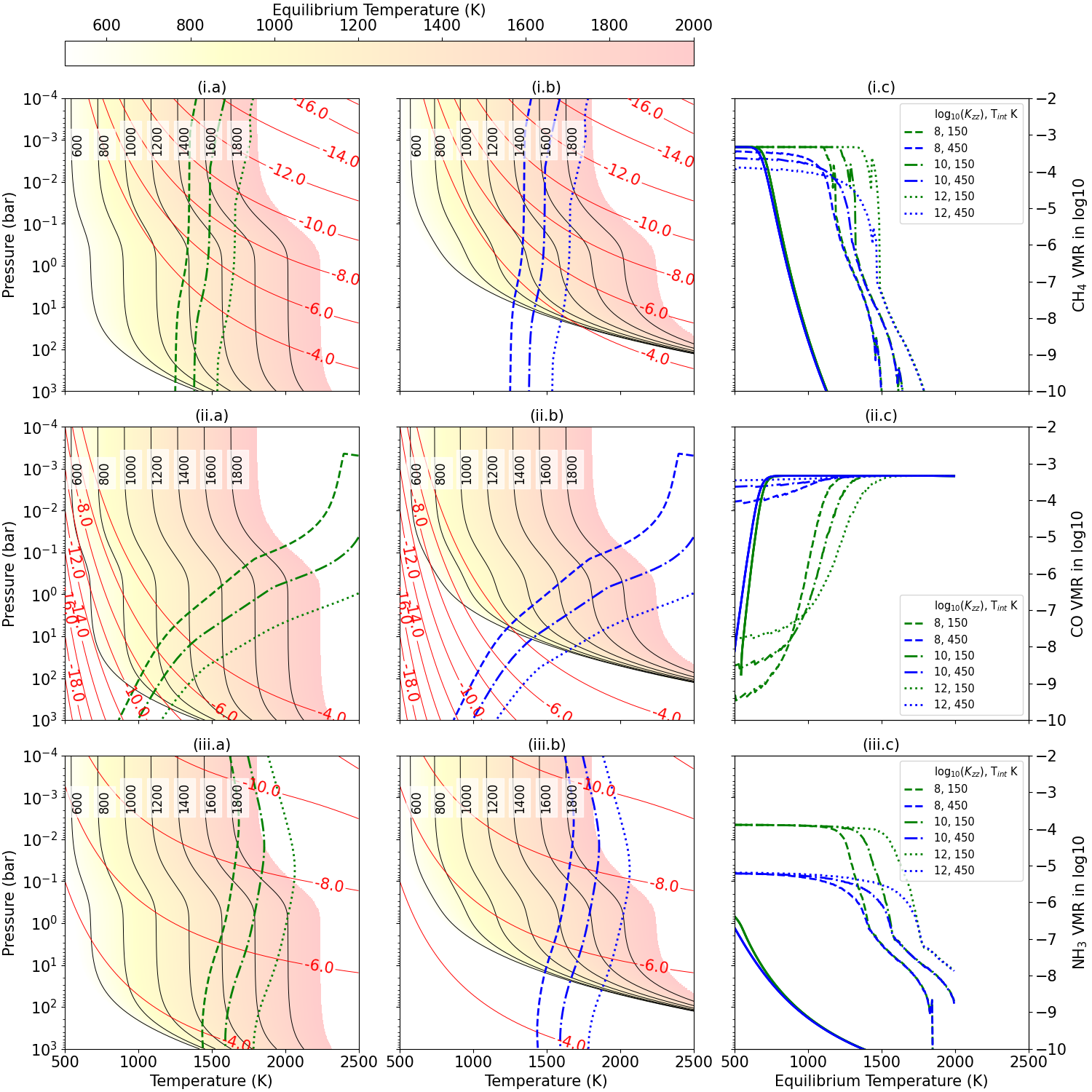}
\caption{Panels (i.c), (ii.c), and (iii.c) show the mixing ratio of \ch{CH4}, \ch{CO}, and \ch{NH3}, respectively, at 1 mbar pressure level and solar metallicity, for a range of equilibrium temperatures. The dashed, dotted-dashed and dotted lines are the mixing ratio for $\log_{10} (K_{zz})$ = 8, 10, and 12, respectively for $T_{\text{int}} = 150$ K (green colored lines) and $T_{\text{int}} = 450$ K (blue colored lines). The solid lines are the thermochemical equilibrium mixing ratio. In panel (i.a), the solid black lines are the thermal profile for a range of equilibrium temperatures and $T_{\text{int}} = 150$ K, while the solid red lines are the contours of the thermochemical equilibrium abundance of \ch{CH4}. The dashed, dotted-dashed and dotted lines are the quenched curves for $\log_{10} (K_{zz})$ = 8, 10, and 12, respectively (see Figure 9 of \citealt{Soni2023a} for quenched curves). Panel (i.b): same as panel (i.a) but for $T_{\text{int}} = 450$ K. Panels (ii.a) and (ii.b): same as panels (i.a) and (i.b), respectively, but for \ch{CO}. Panels (iii.a) and (iii.b): same as panels (i.a) and (i.b), respectively, but for \ch{NH3}.\label{fig:1}}
\end{figure}

\subsection{Effect on the Abundance}\label{subsec:abun}
Panels (i.c), (ii.c), and (iii.c) of Figure \ref{fig:1} show the steady-state abundance of \ch{CH4}, \ch{CO}, and \ch{NH3} at 1 mbar pressure level and solar metallicity for equilibrium temperature $T_{\text{equi}}$ in the range of $500 - 2000$ K, internal temperature $T_{\text{int}}$ = 150 K and 450 K, and strength of transport $\log_{10} (K_{zz})$ = 8, 10, 12. The thermochemical equilibrium abundance of these species is also shown. The figure also shows the contours of the mixing ratio of \ch{CH4}, CO, and  \ch{NH3} over-plotted with the thermal profiles for various equilibrium temperatures (panels (i.a), (i.b), (ii.a), (ii.b), (iii.a), (iii.b)), and the three quenched curves that are shown correspond to $\log_{10} (K_{zz})$ = 8, 10, and 12. 

In the case of CO, for high internal temperature ($T_{\text{int}} = 450$ K, $T_{\text{equi}}>500$ K), the CO quench level lies in the CO-dominant region. Although CO quenches in the atmosphere, vertical mixing does not affect its abundance at 1 mbar. CO follows its thermochemical equilibrium abundance for $T_{\text{equi}}>1000$ K. For these thermal profiles, the planet's quench level of CO and transit infrared photosphere lies in the CO-dominant region; for $T_{\text{equi}}<1000$ K, the photosphere part of the thermal profile enters the \ch{CH4}-dominant region, and the thermochemical equilibrium abundance of \ch{CO} quickly decreases. However, CO remains the dominant C-bearing species at its quench level, making it a major C-bearing species in the infrared photosphere. In the case of low internal temperature ({$T_{\text{int}}$ = 150 K}), the quench level of CO moves into the \ch{CH4}-dominant region as the equilibrium temperature decreases from 1400 K to 500 K. As a result, the CO mixing ratio in the presence of vertical mixing decreases with decreasing equilibrium temperature. However, this transition from CO to \ch{CH4} region occurs for a higher equilibrium temperature in high-pressure regions compared to low-pressure regions. For example, at 100 bar pressure, this transition takes place around 1600 K equilibrium temperature, whereas at 1 mbar pressure, this transition takes place around 800 K equilibrium temperature  \citep{Lodders2002}. As a result, CO quenches in the \ch{CH4}-dominant region even if \ch{CO} remains the dominant C-bearing species in the infrared photosphere in the thermochemical equilibrium composition. Thus, vertical mixing decreases the CO abundance.

\ch{CH4} is the major C-bearing species in the low-temperature and high-pressure regions  \citep{Lodders2002}. For low internal temperature ({$T_{\text{int}} = 150$ K}) and low equilibrium temperatures ($T_{\text{equi}} < 800$ K), the thermal profile remains in the \ch{CH4}-dominant region. As a result, vertical mixing does not affect the \ch{CH4} abundance. The increase in the equilibrium temperature causes the thermal profile to shift into the CO-dominant region. The equilibrium temperature of this transition depends upon the pressure level; at a lower pressure level, this transition occurs at a lower equilibrium temperature (for 0.1 bar pressure, the transition equilibrium temperature is 1000 K). For higher transport strength, the \ch{CH4} quench level shifts into the high-pressure region, and thus, it can remain in the \ch{CH4}-dominant region for a higher value of the equilibrium temperature. The steep nature of the quenched curve of \ch{CH4} causes a sharp decrease in the \ch{CH4} quenched abundance with increasing equilibrium temperature after this transition. The thermochemical equilibrium abundance of \ch{CH4} remains constant for $T_{\text{equi}} <$ 700 K at one mbar and decreases with increasing equilibrium temperature for $T_{\text{equi}} > 700$ K. For very high equilibrium temperatures ($T_{\text{equi}} > 1700$ K), the quench level of \ch{CH4} lies above the one mbar pressure, and quenched \ch{CH4} follows its thermochemical equilibrium abundance.

The equal abundance curve of \ch{NH3}/\ch{N2} lies in the lower temperature and higher pressure region compared to the \ch{CH4}/\ch{CO} equal abundance curve  \citep{Lodders2002}. In the our parameter range, the thermal profile never lies completely in the \ch{NH3}-dominant region, and vertical mixing affects the \ch{NH3} mixing ratio except at a very high equilibrium temperature and low $K_{zz}$ ($\log_{10}(K_{zz})<8$ and $T_{\text{equi}}>$ 1800 K) where the quench level of \ch{NH3} lies above the infrared photosphere of the planet. The contour of the \ch{NH3} mixing ratio follows the thermal profile in the deep atmosphere where the quench level of \ch{NH3} lies. As a result, the change of transport strength changes the quench pressure, but the quench level remains on the same \ch{NH3} contour. For high equilibrium temperature ($T_{\text{equi}}>$ 1000 K), the thermal profile does not follow the contour of \ch{NH3}, leading to a change in the \ch{NH3} mixing ratio with a change in $K_{zz}$. Since the equal abundance curve of \ch{NH3}/\ch{N2} lies in a higher pressure region compared to the \ch{CH4}/\ch{CO} equal abundance curve  \citep{Lodders2002}, the internal temperature significantly affects the quenched abundance of \ch{NH3}. High internal temperatures shift the quench level of \ch{NH3} to a relatively lower pressure region (as evident from Figure \ref{fig:1}). The mixing ratios of \ch{CH4} and \ch{NH3} both increase with pressure. However, the \ch{CH4} mixing ratio saturates at its maximum value once the thermal profile enters the \ch{CH4}-dominant region, and reducing the internal temperature does not change the quenched \ch{CH4} mixing ratio. In contrast, the thermal profile enters the \ch{NH3}-dominant region at a lower internal temperature compared to the \ch{CH4}-dominant region. Consequently, the \ch{NH3} abundance can serve as a good tracer of the internal temperature, which is discussed in Section \ref{subsec:spectra}. 

Thus, we see that the effects of vertical mixing on the abundance of \ch{CH4}, CO and \ch{NH3} are different in different parameter spaces. Vertical mixing has a large effect on the \ch{NH3} mixing ratio, though for a fixed internal temperature, the \ch{NH3} mixing ratio is independent of $K_{zz}$ and $T_{\text{equi}}$ up to a certain value of the equilibrium temperature and after this value, it sharply decreases with $K_{zz}$. For \ch{NH3}, all three values of $K_{zz}$ ($\log_{10} (K_{zz})$ = 8, 10, and 12) give almost the same mixing ratio for lower equilibrium temperature ($T_{\text{equi}}<1200$ K). For low internal temperature ($T_{\text{int}}$ = 150 K), the \ch{CH4} mixing ratio is unaffected by the change in the vertical mixing at low equilibrium temperature and shows a similar behaviour as \ch{NH3}. 

\subsection{Signature on the Transit Spectra}\label{subsec:spectra}
The output spectra of an exoplanet depend on its underlying chemical composition and thermal profile. As discussed in the previous section, the effect of vertical mixing on molecular abundances is molecule-dependent. This effect is more pronounced in some molecules and individually depends on several parameters, including equilibrium temperature, internal temperature, and elemental composition. The signatures of vertical mixing on the transmission spectra are directly link to how significantly vertical mixing affects the atmospheric composition.  Thus there is a narrow range of atmospheric parameters for which vertical mixing has a large affect on the atmospheric transmission spectrum (also termed as the sweet spot) due to transport-induced quenching from a region of the atmosphere where there are large shifts in the quenched abundances of \ch{CH4}-CO or \ch{NH3}-\ch{N2}. This narrow range of atmospheric parameters have been previously studied by \cite{Kawashima2021}, \cite{Zamyatina2023}, and \cite{Sing2024Natur}.

For solar elemental abundance (O/H = $6.06 \times 10^{-4}$, C/H = $2.77 \times 10^{-4}$, and N/H = $8.18 \times 10^{-5}$), oxygen is more than twice as abundant as carbon. This makes the \ch{H2O}  the major reservoir of oxygen, and its mixing ratio remains close to $3\times10^{-4}$ throughout the atmosphere. As a result, it is not affected by vertical mixing. The transit spectrum of the planet is overwhelmed by the \ch{H2O} signature due to its high mixing ratio and large cross section \citep{Madhusudhan2016}. To see the signature of other molecules, the abundance of molecules of interest should be large enough to overcome the \ch{H2O} contribution and also depend upon the noise floor of the observation. The detection of a molecule also depends upon the wavelength region of interest and spectral resolution.  \cite{MacDonald2017} studied the excess due to the \ch{NH3} in the transit spectrum and found that a 100 ppm transit excess due to \ch{NH3} can be obtained by an \ch{NH3} to \ch{H2O} ratio of around $10^{-2}$. Recently, \cite{Gasman2022} studied the minimum abundance of hydrocarbon molecules to be in the detectable level by JWST and concluded that a mixing ratio of $10^{-6} - 10^{-7}$ is required to see the signature of \ch{CH4}. It is evident from Figure \ref{fig:1} that the abundance of \ch{NH3} and \ch{CH4} are sensitive to $K_{zz}$ in the high equilibrium temperature region. However, the mixing ratio of these species is low; therefore, it is not feasible to put any constraint on their abundance (mixing ratio of $10^{-6}$). Thus, the signature of vertical mixing in the transit spectrum is negligible for high equilibrium temperatures. 

\begin{figure}[t!]
	\centering
	\includegraphics[width=1\textwidth]{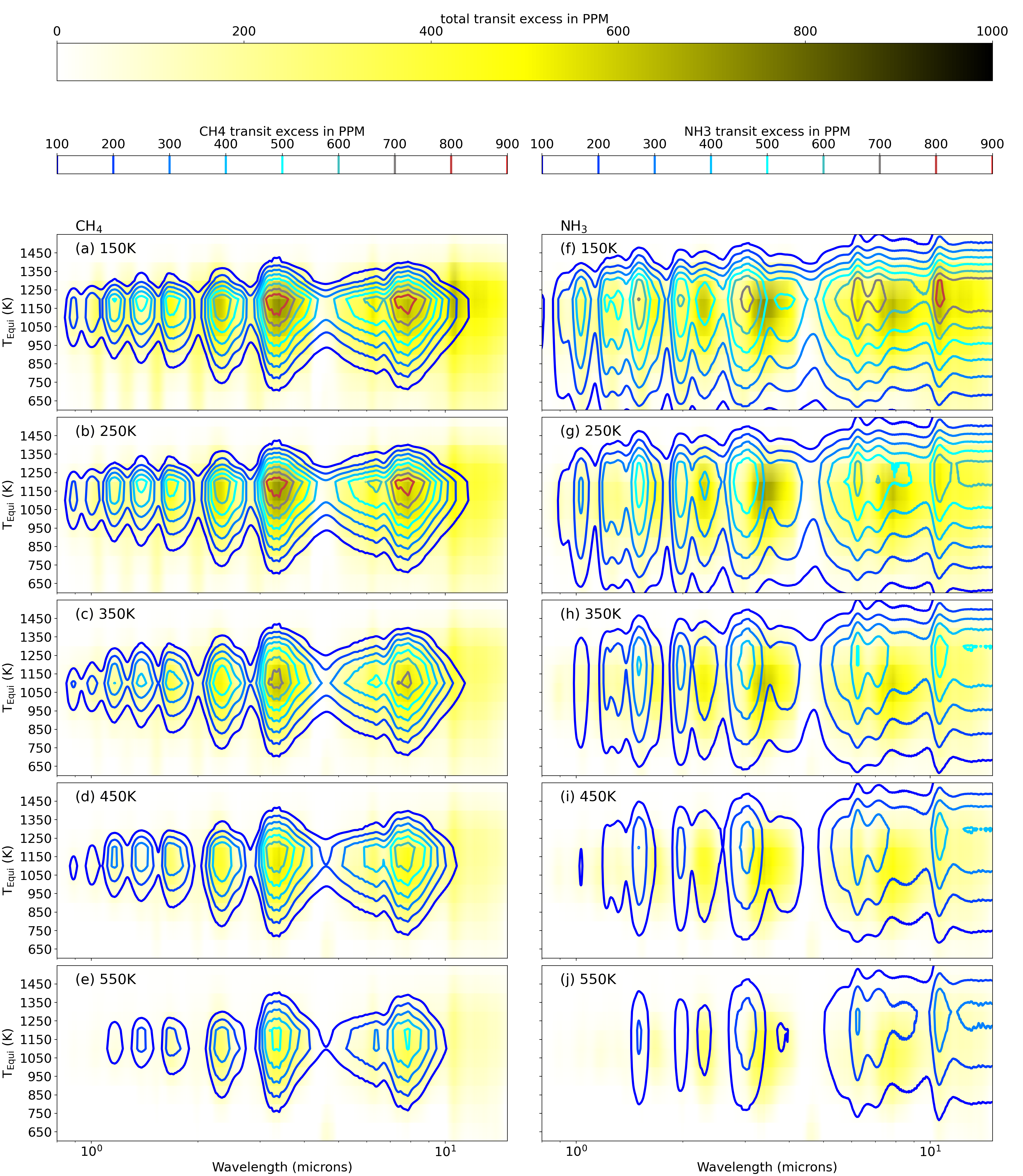}
	\caption{\label{fig:2} The transit depth excess due to vertical mixing in the exoplanet atmosphere from our 1D chemical kinetics model run ($T_{\text{equi}}$ = $500 - 2000$ K; $T_{\text{int}}$ = 150, 250, 350, 450 and 550 K; $\log_{10} (K_{zz})$ = 9; surface gravity = 1000 cm s$^{-2}$). The color contours from blue to red in panels (a) - (e) represent the transit depth excess due to \ch{CH4} and panels (f) - (j) due to \ch{NH3}. The white to black colormesh plot in all the panels represent the total transit depth excess.}
\end{figure}

In Figure \ref{fig:2}, we present the excess in the transit spectrum resulting from vertical mixing across a range of equilibrium and internal temperatures. The transit spectrum was generated using petitRADTRANS, based on outputs from both a chemical equilibrium model and a 1D chemical kinetics model. The figure illustrates the difference between the transit spectra from these two models. 

It is apparent from Figure \ref{fig:2} that the excess due to vertical mixing is largely affected by the equilibrium and internal temperatures. The \ch{NH3} spectrum shows the signature of vertical mixing in a larger range of equilibrium temperature as compared to the \ch{CH4} (\ch{NH3}: $500-1450$ K, \ch{CH4}: $750 - 1350$ K). The contribution of \ch{CH4} in the excess of transit signature due to vertical mixing becomes negligible in the lower equilibrium temperature $<750$ K. However, the contribution of \ch{NH3} remains at the detectable level. As discussed in Section \ref{subsec:abun}, the \ch{NH3} signature is more sensitive to the internal temperature than \ch{CH4}. The maximum excess in the wavelength and equilibrium temperature parameter space due to \ch{NH3} decreases from 800 ppm to 400 ppm when the internal temperature increases from 150 K to 450 K, whereas for \ch{CH4}, this excess decreases from 800 ppm to 600 ppm. The main reason for this is that the \ch{NH3}/\ch{N2} equal abundance curve is located in a deeper part of the atmosphere as compared to the \ch{CH4}/\ch{CO} boundary.

For all the internal temperatures, on increasing the equilibrium temperature, the signature of vertical mixing increases and becomes maximum around 1150 K. Increasing the equilibrium temperature further decreases the signature very quickly, and for a high equilibrium temperature ($>$ 1500 K), the signature of vertical mixing becomes negligible. On increasing the equilibrium temperature, the \ch{CH4} signature decreases more rapidly compared to the \ch{NH3} signature because the \ch{CH4} abundance decrease more steeply as compared to the \ch{NH3} abundance with increasing equilibrium temperature (Figure \ref{fig:1}).

A change in $K_{zz}$ will change the quenched abundance of the major tracers of vertical mixing (\ch{CH4} and \ch{NH3}). As shown in Figure \ref{fig:1}, for the value of $\log_{10} (K_{zz})$ = 8, the \ch{NH3} and \ch{CH4} abundance reach their lowest detectable abundance at a lower equilibrium temperature ($\approx$ 1250 K for \ch{CH4} and $\approx$ 1450 K for \ch{NH3}) compared to the higher value of $K_{zz}$ ($\log_{10} (K_{zz})$ = 12). As a result, the signature of vertical mixing in the transit spectrum is extended towards high equilibrium temperature values with increasing eddy diffusion coefficient. The high $K_{zz}$ will shift the quench level deeper in the atmosphere, increasing the \ch{NH3} and \ch{CH4} quenched abundance  \citep{Lodders2002}. 

The abundance of \ch{H2O} increases linearly with metallicity in the entire parameter space, whereas for CO, \ch{CH4} and \ch{NH3}, the abundance increases linearly with metallicity in their respective dominant regions.  Metallicity has a minimal effect on the \ch{CH4} abundance in the CO-dominated region \citep{Lodders2002, Soni2023a}. The \ch{CH4}/\ch{CO} and \ch{NH3}/\ch{N2} equal abundance curves shift toward the higher pressure and lower temperature region with increasing metallicity \cite{Lodders2002}. For the solar metallicity, the planets whose thermal profiles lie entirely in the \ch{CH4}-dominant region or whose \ch{CH4} quench levels lie in the \ch{CH4}-dominant region (lower equilibrium temperature planets), an increase in metallicity changes the deeper part of the atmosphere from \ch{CH4}-dominant to CO-dominant  \citep{Lodders2002, Visscher2011, Soni2023a}. As a result, the quench level of \ch{CH4} also shifts from a \ch{CH4}-dominant to a CO-dominant region. The \ch{CH4} abundance does not increase with metallicity in the region where  CO  is dominant. However, the \ch{H2O} signature strength increases, which leads to \ch{CH4} contributing less to the overall transit spectrum.

\section{Retrieval Results} \label{sec:rslt}

\begin{table}

\begin{center}
\renewcommand{\arraystretch}{1.3}
\caption{Parameters of the chemical kinetics model runs}
\label{Table5.1a}
\begin{tabular}{ |c|c|c|c| } 
\hline
Parameter & Initial value & End value & Step size \\ 
\hline
Equilibrium temperature $T_{\text{equi}}$ (K) & 500  & 2000 & 100\\ 
\hline
Internal temperature $T_{\text{int}}$ (K)  & 100 & 550 & 50\\ 
\hline 
Surface gravity $\log_{10} (g)$ (cm s$^{-2}$)  & 2  & 5 & 0.5\\ 
\hline 
Atmospheric metallicity [M/H] & -1 & 2 & 0.5\\ 
\hline 
Radius of the planet ($R_J$) & 1  & 1 & fixed\\ 
\hline
Reference pressure (bar)& 0.1 & 0.1  & fixed\\ 
\hline
Eddy diffusion coefficient $\log_{10} (K_{zz})$ (cm$^2$ s$^{-1}$)& 4 & 12 & 2  \\ 
\hline	
\end{tabular}
\end{center}
\end{table}

\begin{figure}[t!]
	\centering
	\includegraphics[width=1\textwidth]{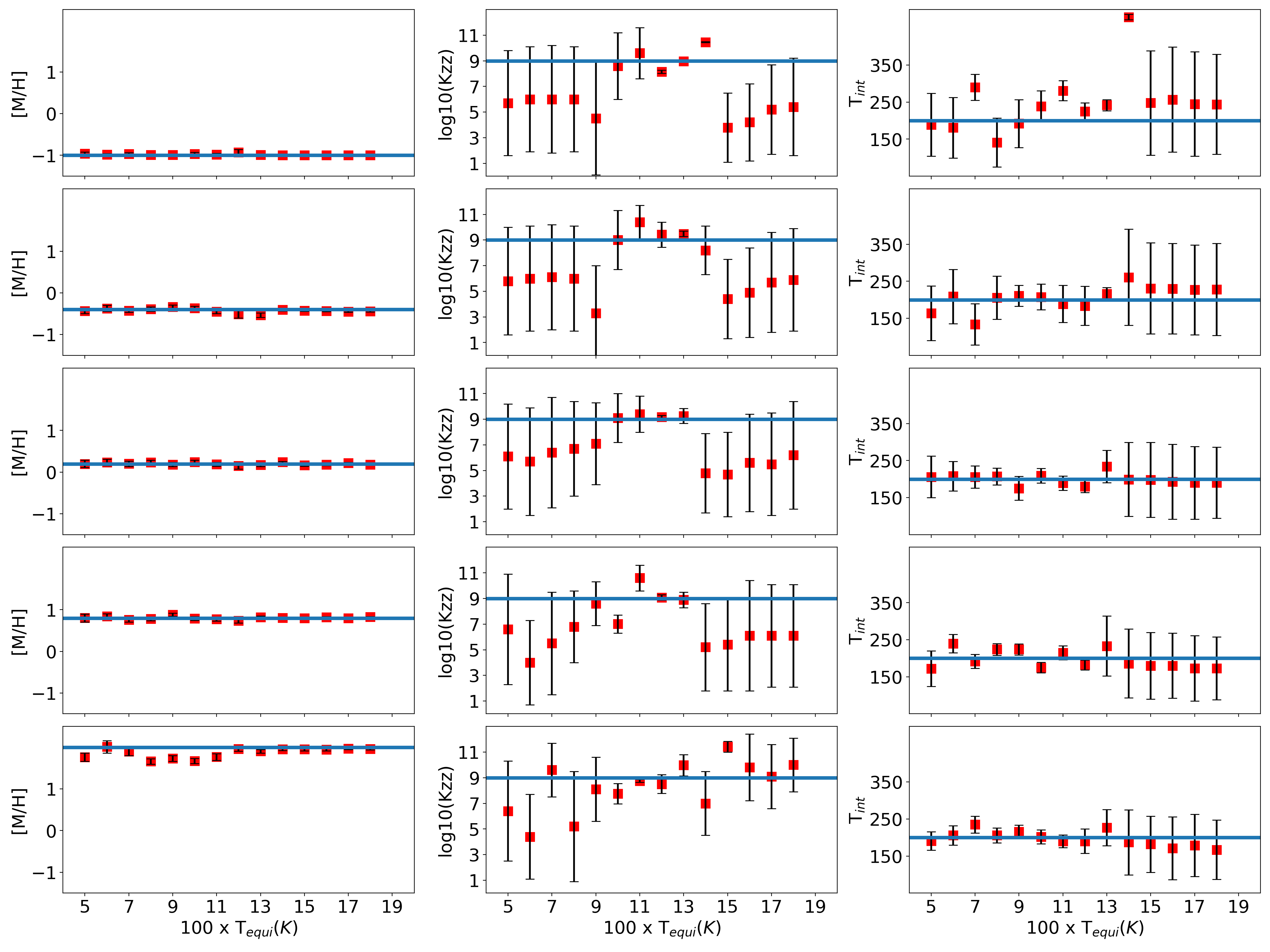}
	\caption{The retrieved values of [M/H], $K_{zz}$ and $T_{\text{int}}$ for different values of metallicity and equilibrium temperature with $\log_{10} (K_{zz}) = 9$, $T_{\text{int}} = 200$ K and $g = 1000$ cm s$^{-2}$. The rows from top to bottom correspond to metallicity values of -1, -0.4, 0.2, 0.8, and 2. The blue horizontal lines show the true values of the parameters. \label{fig:met}}
\end{figure}

\begin{figure}[t!]
	\centering
	\includegraphics[width=1\textwidth]{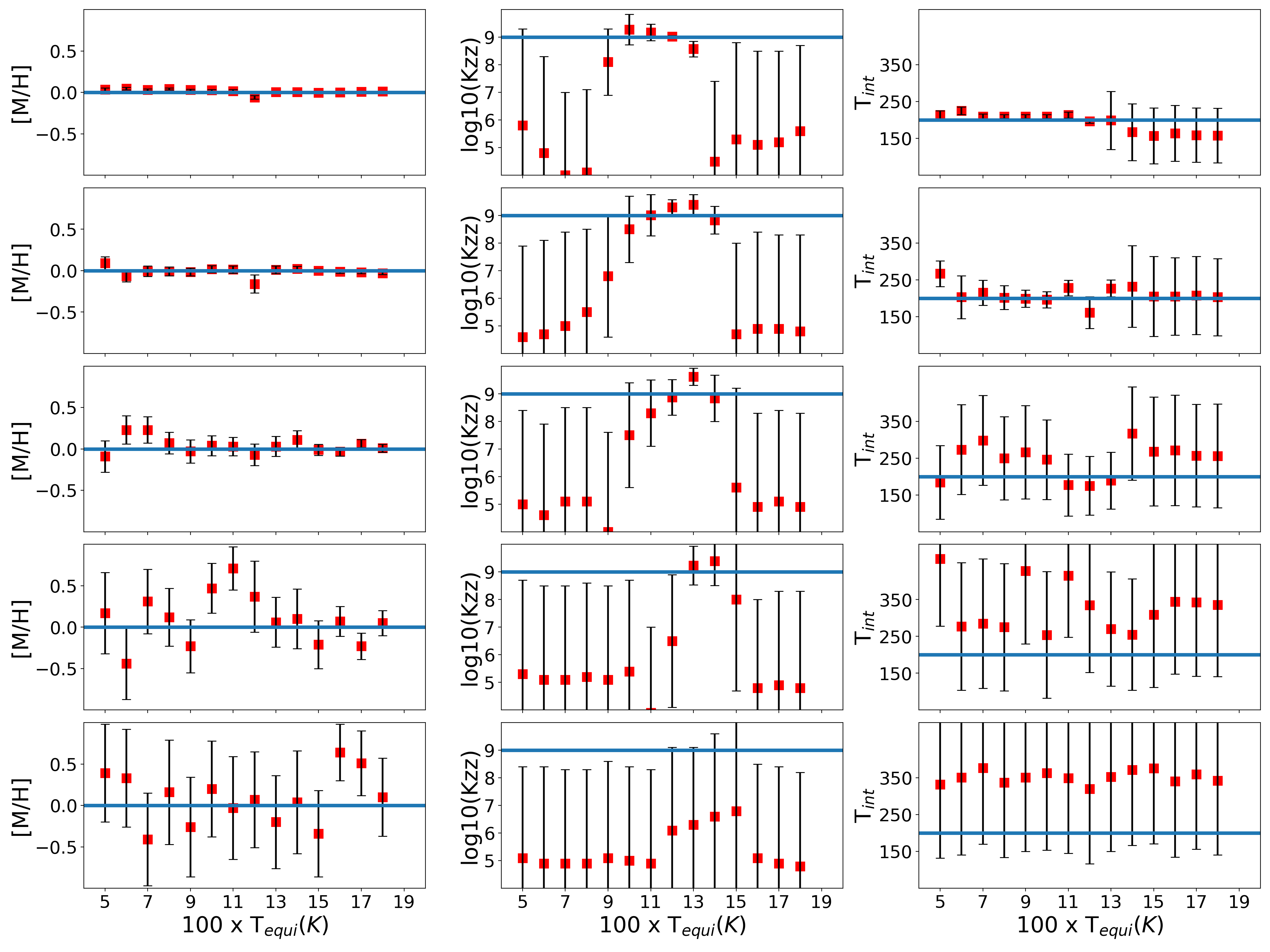}
	\caption{The retrieved values of [M/H], $K_{zz}$ and $T_{\text{int}}$ for different values of surface gravity and equilibrium temperature with $\log_{10} (K_{zz}) = 9$, $T_{\text{int}} = 200$ K and solar metallicity. The rows from top to bottom correspond to surface gravity values of 3, 10, 30, 100, and 300 m s$^{-2}$. The blue horizontal lines show the true values of the parameters. \label{fig:grav}}
\end{figure}

\begin{figure}[t!]
	\centering
	\includegraphics[width=1\textwidth]{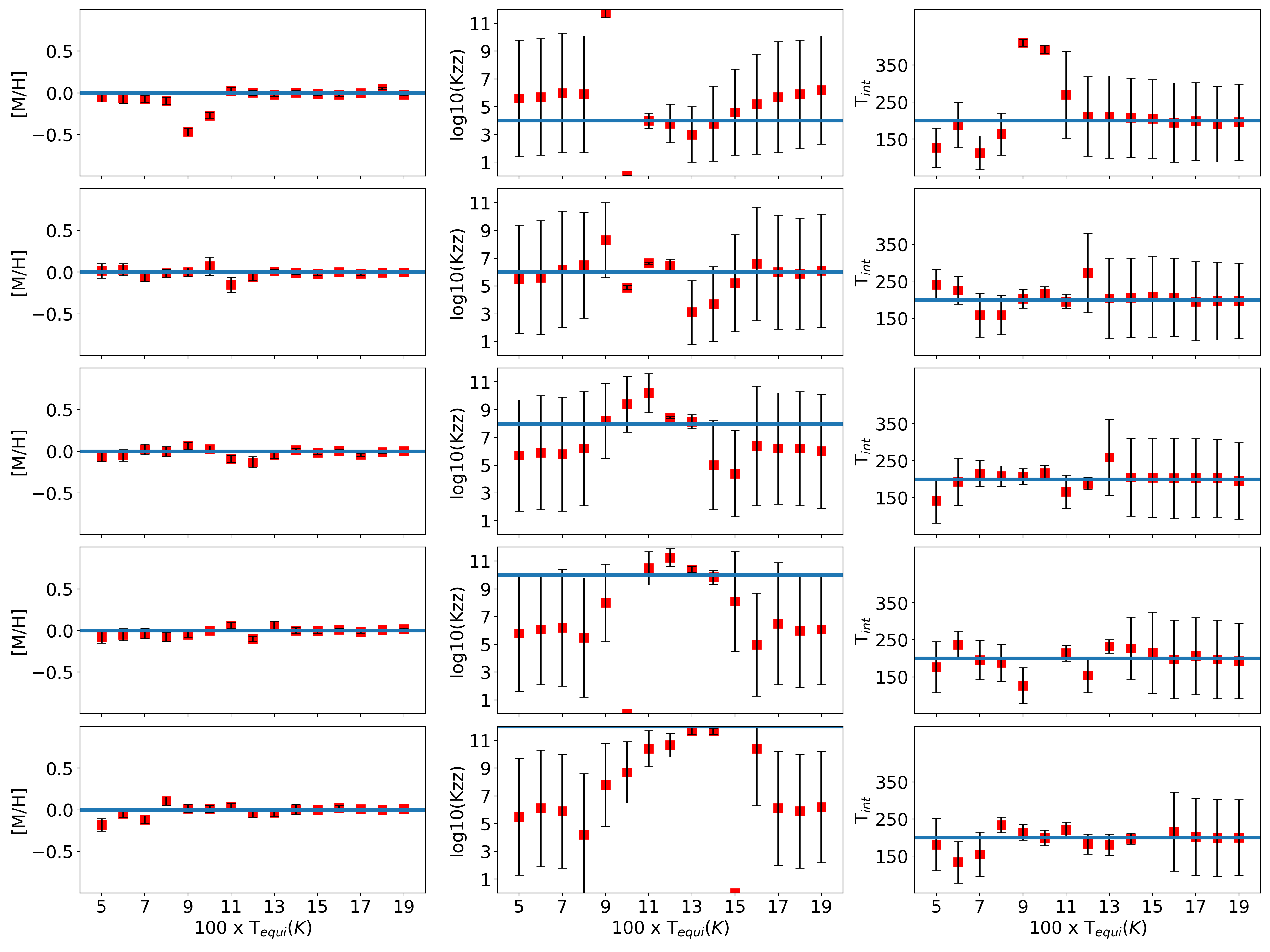}
	\caption{The retrieved values of [M/H], $K_{zz}$ and $T_{\text{int}}$ for different values of eddy diffusion coefficient and equilibrium temperature with $T_{\text{int}} = 200$ K, $g = 1000$ cm s$^{-2}$ and solar metallicity. The rows from top to bottom correspond to $\log_{10} (K_{zz})$ values of 4, 6, 8, 10, and 12. The blue horizontal lines show the true values of the parameters. \label{fig:kzz}}
\end{figure}

We ran more than 300 chemical kinetics models, which span the parameter space of equilibrium temperature, internal temperature, atmospheric metallicity, transport strength and surface gravity. Table~\ref{Table5.1a} shows how our model parameters are distributed in the parameter space. We generated the JWST synthetic spectrum for each model run and ran the retrieval model. We analyzed the retrieved values of $K_{zz}$, metallicity and internal temperature for all our retrieval model outputs. The statistical constraints on these parameters in the retrieval output are directly linked to the inherent signature of vertical mixing in the transit spectrum. Figures \ref{fig:met} to \ref{fig:kzz} show the retrieved parameters and the true values for the parameters listed in Table~\ref{Table5.1a}.

Figure \ref{fig:met} shows the retrieval output of the models run in the parameter space of metallicity and equilibrium temperature. The metallicity  strongly affects the transit spectrum of the planet, thus the statistical constraint on the retrieved metallicity is high compared to $K_{zz}$ and internal temperature. For most of the models, the retrieved value of metallicity is very close to its true value, and for higher metallicity models, the retrieved value is within 10\% of its true value. The retrieval constraints on the internal temperature mainly follow the \ch{NH3} abundance profile (see previous section). For these model runs, the true value of $\log_{10} (K_{zz}) = 9$, and in Figure \ref{fig:1}, for $\log_{10} (K_{zz})$ = 8 and 10, the \ch{NH3} abundance starts to decrease after 1100 K and the mixing ratio reaches below the detectable level at 1400 K. For high equilibrium temperature, the \ch{NH3} quench pressure does not change with the internal temperature, as the quench level lies in the part of the thermal profile (pressure $<$ 1 bar) where the thermal profile becomes independent of the internal temperature. As a result, the quenched \ch{NH3} abundance does not change with the internal temperature (see Figure B1 in \citealt{Soni2023b}). The retrieved internal temperature is tightly constrained for an equilibrium temperature below 1400 K and loosely constrained for a higher equilibrium temperature. Increasing the metallicity also helps to put a tighter constraint on the internal temperature. Increasing the metallicity increases the \ch{NH3} abundance, and the \ch{CH4}/CO boundary moves deep in the atmosphere. This deep \ch{CH4}/CO boundary makes the \ch{CH4} abundance more sensitive to vertical mixing in the high-pressure region. 

The constraints on the transport strength ($K_{zz}$) highly depend upon the equilibrium temperature; for lower equilibrium temperature ($<$ 900 K), the $K_{zz}$ values are loosely constrained even if the transit spectrum has a large signature of vertical mixing (see Figure \ref{fig:2}). As described in the previous section, the change in $K_{zz}$ does not affect the \ch{NH3} and \ch{CH4} abundance for lower equilibrium temperatures. For high equilibrium temperatures,  \ch{CH4} and \ch{NH3} are not present in the detectable level, and CO is not affected by the vertical mixing, resulting in $K_{zz}$ being loosely constrained in this temperature region. For moderate equilibrium temperatures ($900$ K $< T_{\text{equi}} < 1400$ K), the degeneracy in the $K_{zz}$ parameter space is small, and the retrieval models perform better to constrain $K_{zz}$. However, this equilibrium temperature range inherently depends upon the physics involved to calculate the thermal profile and the atmospheric parameters, including elemental abundance and internal temperature. The change in metallicity also alters this temperature range, though its effect can be complex, as the \ch{CH4}/\ch{CO} and \ch{NH3}/\ch{N2} boundaries move deep in the atmosphere with an increase in metallicity. Also, these boundaries become steeper with pressure \citep{Soni2023a, Soni2023b}, thereby reducing the range of equilibrium temperature thermal profiles that cross these boundaries. On the other hand, an increase in metallicity increases the \ch{NH3} abundance, leading to an increase in its detectability for higher equilibrium temperatures. 

In Figure \ref{fig:grav}, we have plotted the retrieval output of the models run in the parameter space of surface gravity and equilibrium temperature. The gravity changes the shape of the thermal profile and the scale height of the planet. The output transit spectrum significantly depends upon the scale height since the amplitude of the transit feature in the spectrum is directly proportional to the scale height \citep{Seager2010, Kreidberg2018b}. This is also reflected in the retrieval output. With increasing gravity, the statistical uncertainty in the retrieved parameters increases. For gravity $> 300$ m s$^{-2}$, the uncertainty covers the prior parameter range. 

In Figure \ref{fig:kzz}, we present the retrieval outputs of the models run within the parameter space of the eddy diffusion coefficient $K_{zz}$ and equilibrium temperature. $K_{zz}$ influences the quench level of the chemical species, thereby affecting the abundances of \ch{CH4} and \ch{NH3}. An increase in $K_{zz}$ pushes the quench level of \ch{NH3} deeper into the atmosphere, corresponding to regions of higher pressure. For a given equilibrium temperature, variations in the internal temperature primarily impact the thermal profile in these high-pressure regions. Consequently, with higher $K_{zz}$ values, \ch{NH3} is quenched in parts of the atmosphere influenced by the internal temperature. As a result, a higher value of $K_{zz}$ can provide constraints on the internal temperature, particularly for planets with higher equilibrium temperatures. For instance, at $\log_{10} (K_{zz})$ = 4, the internal temperature is well-constrained for $T_{\text{equi}} < $ 1100 K, while at $\log_{10} (K_{zz})$ = 12, it is well-constrained for $T_{\text{equi}} < $ 1600 K.

\section{Discussion}\label{sec:discuss}
The effect of vertical mixing on the atmospheric composition largely depends upon the thermal profile of the planet. In this study, we have used a parametric thermal profile from \cite{Guillot2010}, in which the thermal profile depends upon the equilibrium temperature, the internal temperature, the infrared opacity $\kappa_{ir}$ = 0.01, and the ratio of the visible to infrared opacity $\gamma$ = 0.4. 
Using different values of $\kappa_{ir}$ or $\gamma$ may change  the range of the parameters for the signature of the vertical mixing. However, the major conclusions of this study will remain unchanged.

Vertical and horizontal mixing affect the exoplanet atmospheric composition due to zonal advection fast winds. However, this study focuses on the effect of vertical mixing and its signature in the observed spectrum through running the 1D disequilibrium models. Some recent studies have included both vertical and horizontal advection mixing. \cite{Baeyens2021} used 3D modeling and concluded that below an effective temperature of 1400 K, the atmosphere is homogeneous and vertical mixing governs the atmospheric composition. Above 1400 K, the atmosphere shows longitudinal diversity in the chemical composition. \cite{Zamyatina2023} studied the effect of transport-induced quenching by running a 3D cloud and haze-free model for the atmospheres of four gas giants. They also concluded that there is a sweet spot in the planetary parameter space for which the signature of transport-induced quenching is maximum in the transit spectrum. The temperature should be large enough to have a short advection timescale and small enough to have a large chemical timescale. Our results agree with these 3D modeling studies even if we include only 1D modeling. This is because the temperature range over which the transport processes show maximum signatures is the same for which vertical mixing is the major process that affects the atmospheric composition. However, further studies are required to understand the signature of vertical mixing in 3D modeling. 

Photochemistry is an important disequilibrium process that largely affects the atmospheric composition above the 1 mbar pressure region. The major vertical mixing tracers (\ch{CH4} and \ch{NH3}) are highly photoactive, and their abundance significantly decreases due to photodissociation \cite{Moses2011, Hu2012}. Typically, the photochemical region lies very high in the atmosphere ($P<10^{-3}$ bar) and photochemistry has a limited effect on the transmission spectrum. 

In this study, we have not incorporated the effects of clouds and hazes. Clouds and hazes can suppress the transit spectrum feature, and the signature of vertical mixing may be suppressed in the spectrum. However, the effect of clouds in the transmission spectrum is minimal above 10 $\mu$m. Above 10 $\mu$m, the \ch{NH3} feature is also more prominent \citep{Ohno2023a}. The presence of clouds and hazes also affects the thermal profile of the planet through radiative feedback. The presence of thick clouds and hazes can increase the temperature by increasing the atmosphere's opacity. This increase in temperature around the quench pressure can affect the quenched abundance. On the other hand, clouds and hazes can block the stellar flux reaching deep in the atmosphere, thereby decreasing the temperature \citep{Molaverdikhani2020}. The effect of photochemistry and clouds/hazes are important in understanding the signature of vertical mixing, as both processes can affect the composition of the tracer of vertical mixing in the spectra or can directly affect the planet's spectra. Both effects are out of the scope of this work and will be considered in future studies. 
 
\section{Conclusion}\label{sec:concl}
In this study, we built a fast-forward disequilibrium model which includes vertical mixing using the quenching approximation. This model uses the chemical timescales calculated from our previous work and uses the Smith method to constrain the mixing length of the atmosphere to calculate the vertical mixing timescale of the atmosphere accurately. This model is tested with the open-source atmospheric retrieval code petitRADTRANS and named the quenched level retrieval model (QLR). The model calculated the quench level of major H-C-N-O bearing species (\ch{H2O}, \ch{CO2}, CO, \ch{CH4}, HCN, \ch{NH3} and \ch{N2}). We benchmarked the model with a hierarchy approach by generating a synthetic JWST observation for a test planet atmosphere generated by a 1D chemical kinetics model. The retrieved values are in good agreement with the true values. 

We studied the effect of vertical mixing on the transit spectrum for a large range of parameters ($T_{\text{int}}$, $T_{\text{equi}}$, surface gravity, $K_{zz}$ and atmospheric metallicity) and how well the retrieval model can constrain the strength of vertical mixing from this transit spectrum. We aimed to find the likelihood of constraining $K_{zz}$ in our parameter space using the QLR model. We ran over 300 chemical kinetics models across our parameter space. We generated a JWST synthetic spectrum for each model run and then ran the QLR model. We analyzed the retrieved values of metallicity, $K_{zz}$, and internal temperature. The statistical constraints on these parameters are directly linked to the signatures in the atmospheric spectrum.  This study finds that the $K_{zz}$ can be reasonably constrained only over a relatively narrow range of conditions, given the numerous parameters that connect disequilibrium chemistry to observed spectrum, and any constraint on the $K_{zz}$ should be taken with much caution! This study revealed key findings regarding the impact of transport strength ($K_{zz}$) on the atmospheric composition and its retrieval, summarized as follows.
\begin{itemize}
    \item Low equilibrium temperatures ($T_{\text{equi}} <$ 900 K): constraints on $K_{zz}$ are loose because variations in $K_{zz}$ do not significantly affect the \ch{NH3} and \ch{CH4} abundances, making retrieval models less effective.
    \item Moderate equilibrium temperatures (900 K $<T_{\text{equi}} <$ 1400 K):  constraints on $K_{zz}$  are tighter due to reduced degeneracy and increased sensitivity of \ch{NH3} and \ch{CH4} to vertical mixing, improving retrieval accuracy.  Again, better constraints on $K_{zz}$ can be made when mixing is occurring from a region of the atmosphere where the chemistry is most sensitive to temperature changes.

    \item High equilibrium temperatures ($T_{\text{equi}} >$ 1400 K): \ch{NH3} and \ch{CH4} become undetectable, and CO shows minimal changes, leading to poor $K_{zz}$ constraints due to a lack of detectable vertical mixing signatures.
\end{itemize}

\section*{Acknowledgments}
The authors thank the anonymous referee for their
constructive comments, which strengthened the paper. The computations were performed on the Param Vikram-1000 High Performance Computing Cluster of the Physical Research Laboratory (PRL), India. The work done at PRL is supported by the Department of Space, Government of India. We are thankful to the current and former developers of the petitRADTRANS Python package. We thank Jayesh Goyal and Liton Majumdar for useful insights and discussions. We thank Sana Ahmed for suggestions that improved the overall readability of the manuscript.

\bibliographystyle{aasjournal}
\bibliography{references}

\appendix
\section{Benchmarking} \label{app:benchmaking}
\begin{figure}[t!]
	\centering
	\includegraphics[width=0.45\linewidth]{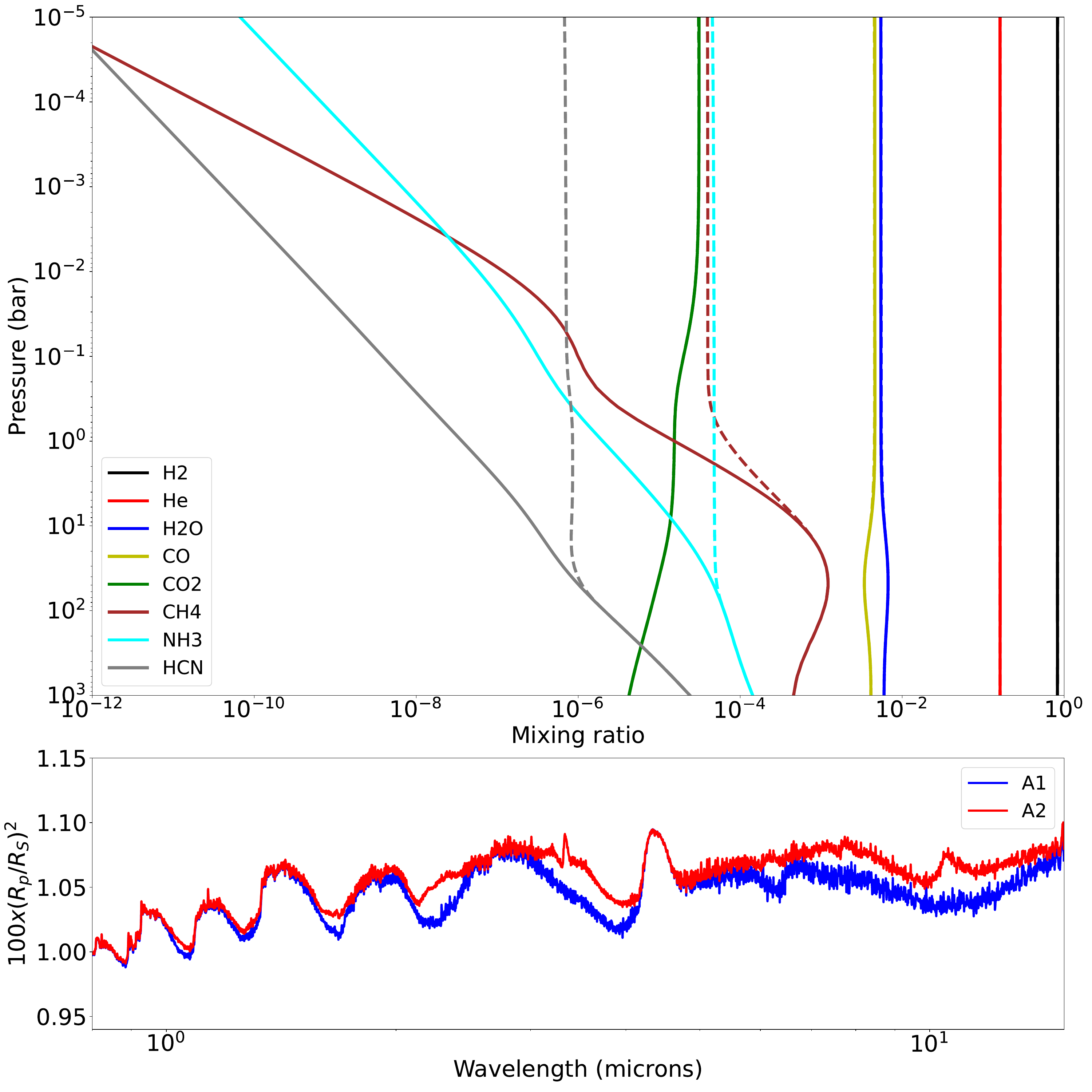}
 	\includegraphics[width=0.45\linewidth]{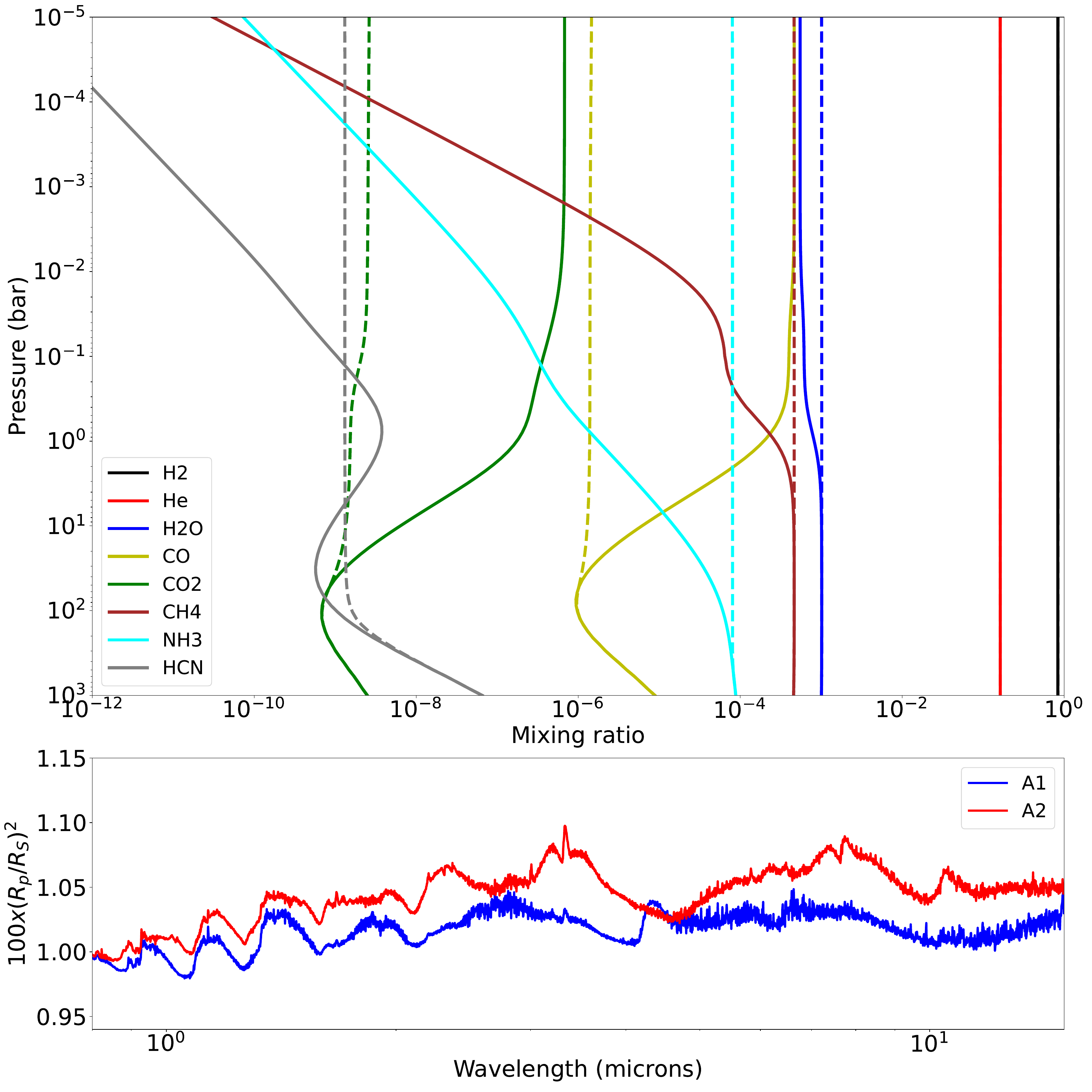}
	\caption{Left column: The top panel shows the vertical volume mixing ratio (VMR) for A1 (thermochemical equilibrium abundance) and A2 (disequilibrium abundance) in solid and dashed lines, respectively. The bottom panel shows the transmission spectra for A1 and A2 in solid blue and red lines, respectively. Right column: similar to the left column but for B1 and B2.}\label{Fig:abundance}
\end{figure}

To benchmark the quench level retrieval (QLR), we generated two test case thermal profiles, A and B, using petitRADTRANS (based on \citealt{Guillot2010}). The parameters for these cases, provided in Table \ref{Table5.1}, were selected such that vertical mixing primarily affects \ch{CH4} and \ch{NH3} abundance in case A, while in case B, the abundances of all major chemical species are influenced. For case A (Figure \ref{Fig:abundance}), the thermal profile remains within the CO-dominant region throughout the atmosphere, making CO the primary carbon-bearing species in the thermochemical equilibrium composition. The CO abundance remains unaffected by vertical mixing, as the quench level of CO lies within the CO-dominated region. This also ensures that \ch{CO2} maintains its thermochemical equilibrium abundance throughout the atmosphere (see Section \ref{sec:quench}). However, vertical mixing significantly increases the abundances of \ch{CH4} and \ch{NH3} by 4–5 orders of magnitude in the infrared photosphere (at pressures around 1 mbar). The quenching of \ch{NH3} also affects the HCN abundance, increasing it by 3–4 orders of magnitude.

\begin{table}[h!]

\begin{center}
\renewcommand{\arraystretch}{1.3}
\caption{Parameters used to calculate the atmospheric composition of the test cases A and B.}
\label{Table5.1}
\begin{tabular}{ |c|c|c|c|c| } 
\hline
Parameter & A1 & A2 & B1 & B2\\ 
\hline
Equilibrium temperature $T_{\text{equi}}$ (K) & 1200  & 1200 & 1000 &  1000\\ 
\hline
Internal temperature $T_{\text{int}}$ (K)  & 300 & 300  & 200 & 200\\ 
\hline 
Surface gravity $g$ (cm s$^{-2}$)  & 10$^3$  &10$^3$& 10$^3$  &10$^3$ \\ 
\hline 
Atmospheric metallicity [M/H] & 1 & 1 & 0 & 0\\ 
\hline 
Radius of the planet ($R_J$) & 1  & 1 & 1  & 1 \\ 
\hline
Reference pressure (bar)& 0.1 & 0.1 & 0.1 & 0.1  \\ 
\hline
Eddy diffusion coefficient $K_{zz}$ (cm$^2$ s$^{-1}$)& 0& $10^9$ & 0 & $10^9$  \\ 
\hline	
\end{tabular}
\end{center}
\end{table}

The lower equilibrium and internal temperatures in case B (Figure \ref{Fig:abundance}) result in a more pronounced variation of atmospheric abundances from their thermochemical equilibrium values. The thermal profile in case B lies in the \ch{CH4}-dominant region at pressures greater than 1 bar, while at pressures less than 1 bar, it transitions to a CO-dominant region. Under thermochemical equilibrium, CO would dominate the infrared photosphere. However, strong vertical mixing ($K_{zz} = 10^9$ cm$^2$ s$^{-1}$) significantly alters the atmosphere, making \ch{CH4} the dominant species over CO by 3 orders of magnitude in the infrared photosphere. The vertical mixing impacts several major HCNO-bearing species, including CO (by approximately 10$^2$), \ch{CO2} (by approximately 10$^2$), \ch{H2O} (by a factor of about 2), \ch{CH4} (by approximately 10$^1$), \ch{NH3} (by approximately 10$^3$), and HCN (by approximately 10$^1$).

We compute the thermochemical equilibrium and disequilibrium mixing ratios using our 1D chemical kinetics model described in \cite{Soni2023a} for each case. Figure \ref{Fig:abundance} shows the equilibrium and disequilibrium atmospheric compositions for cases A and B, along with their corresponding transit spectra. As seen in the figures, the transit spectra of A1 (equilibrium abundance) and B1 (equilibrium abundance) differ from those of A2 (disequilibrium abundance) and B2 (disequilibrium abundance). This spectral difference arises from variations in atmospheric composition. The difference is more pronounced for case B than for case A, which directly reflects the stronger influence of vertical mixing on atmospheric composition in case B.

\subsection{Synthetic Observation}
\begin{figure}[!ht]
	\centering
	\includegraphics[width=1\linewidth]{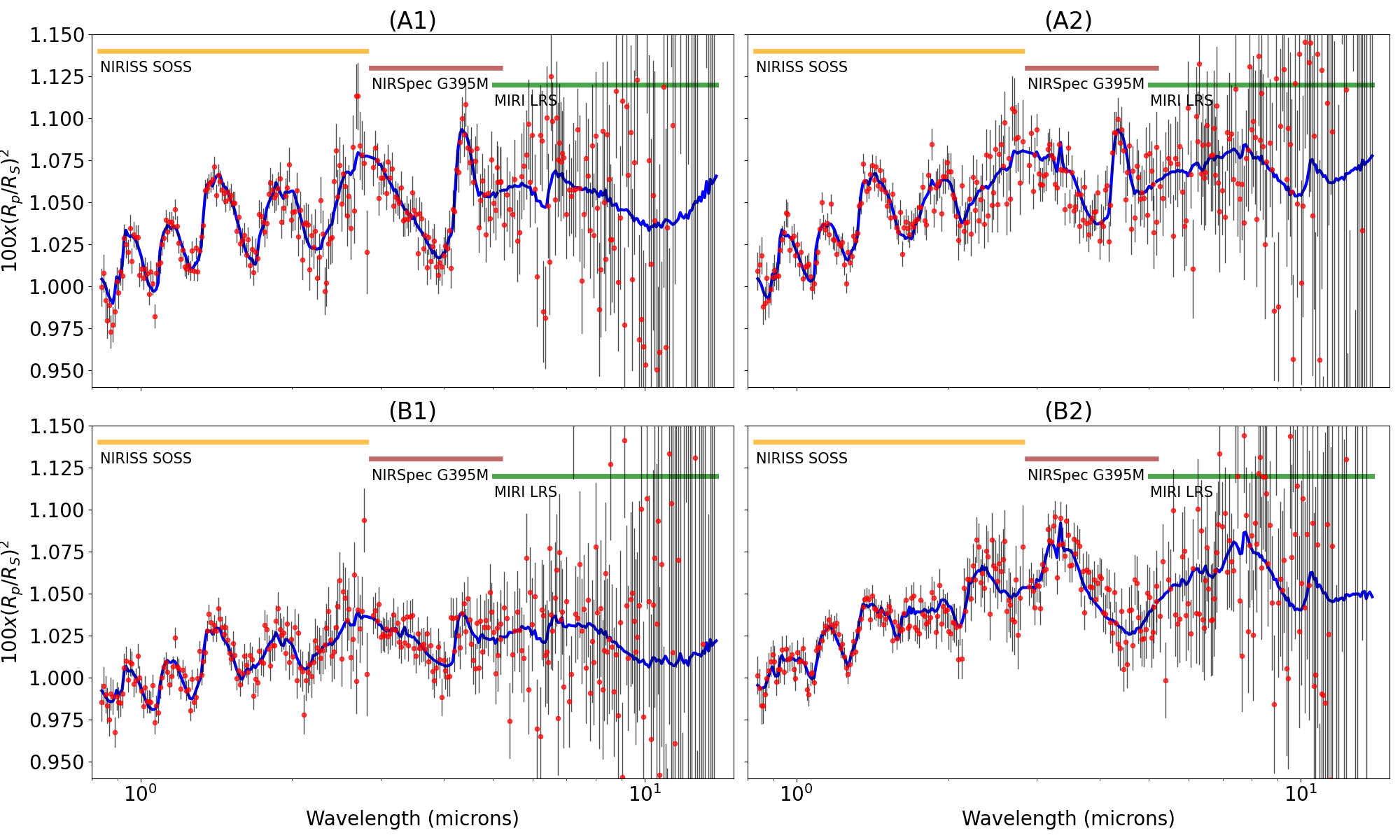}
	\caption{ The solid blue line shows the re-binned transit spectrum of A1 (top left panel), A2 (top right panel), B1 (bottom left panel) and B2 (bottom right panel). The solid red dots show the simulated JWST observed spectrum, and the black vertical lines are the associated errors. The three solid horizontal colored lines (yellow, brown, and green) show the wavelength coverage of the labeled JWST instrument.}\label{Fig:Transmission_data}
\end{figure}

We have generated the synthetic JWST transit spectrum from $0.8 - 14 \mu$m for the two test exoplanets. For this, we used PandExo \citep{Batalha2017}, an open-source Python package, to calculate the instrumental noise for JWST and HST while observing the exoplanet transit. PandExo uses certain Pandeia \citep{Pontoppidan2016} functionality to calculate the exposure time in the observation. Pandeia includes the updated background noise, PSF (point spread function) of the instruments and their optical path, ramp noise, saturation noise, correlated read noise, and flat field error for the JWST instruments. PandExo uses two methods to simulate the noise in the transit spectrum. The first method subtracts the last readout from the first readout frame (in each observation, before the saturation of the instrument, several readouts are made), which is called the LMF method. The second method separately fits each readout frame ( up-the-ramp readout) and calculates the correlated noise. 
The PandExo parameters used for generating the synthetic JWST observation are given in Table \ref{Table5.2}, and the synthetic spectra for the test cases A1, A2, B1 and B2 are shown in Figure \ref{Fig:Transmission_data}.
 We include three JWST instruments (to generate synthetic observation): single object slitless spectroscopy (SOSS) of Near Infrared Imager and Slitless Spectrograph (NIRISS), NIRSpec G395M, and MIRI LRS.

\begin{table}[t!]
	\begin{center}
		\renewcommand{\arraystretch}{1.3}
		\caption{Parameters used to generate the synthetic JWST observation.}
		\label{Table5.2}
		\begin{tabular}{ |c|c| } 
			\hline
			Stellar temperature & 5700 K \\ 
			\hline
			Stellar radius  & 1 solar radius  \\ 
			\hline
			Apparent magnitude  & 10.33 \\ 
			\hline 
			Stellar metallicity  & solar metallicity \\ 
			\hline 
			Stellar gravity  & $10^{4}$ cm s$^{-2}$ \\ 
			\hline 
			Transit duration & 3 hours and 40 minutes  \\ 
			\hline
			$f_{\text{base}}$ & 1  \\ 
			\hline
			Number of transit observations & 2 \\ 
			\hline
			Saturation level & 80\%  \\ 
			\hline
			Noise floor & 20, 75, 40 ppm  \\ 
			\hline
			
		\end{tabular}
	\end{center}
\end{table}

\begin{table}[t!]
\begin{center}
\renewcommand{\arraystretch}{1.3}
\caption{Prior distribution of the retrieval parameters.}
\label{Table5_5}
\begin{tabular}{ |c|c|c|c| } 
\hline
Parameter                                            &  A        & B         & Distribution   \\ 
\hline
$T_{\text{equi}}$ (K)                                & 800-1600  & 800-1600  & evenly         \\ 
\hline
$T_{\text{int}}$ (K)                                 & 50-600   & 50-600     & evenly          \\ 
\hline 
$\log_{10}(K_{zz} / \text{cm}^2 \text{ s}^{-1})$  & -         & 0-12      & evenly          \\ 
\hline
$\log_{10}(g/\text{cm s}^{-2})$                & 1-4       & 1-4       & evenly          \\ 
\hline
[M/H]                                                & -1 to 3   & -1 to 3   & evenly          \\ 
\hline
\end{tabular}
\end{center}
\end{table}

\begin{table}[t!]
\scriptsize
\begin{center}
\renewcommand{\arraystretch}{1.2}
\caption{ Retrieved parameters for the cases A and B.}
\label{Table5_3}
\begin{tabular}{ |c|c|c|c|c|c|c|c|c|c| } 
\hline
Model run  $\rightarrow$  &                        &CCR-A1                &CCR-A2                &QLR-A1                &QLR-A2                 &CCR-B1                &CCR-B2               &QLR-B1               &QLR-B2               \\ 
\hline
Parameters $\downarrow$   &True value $\downarrow$ &                      &                      &                      &                       &                      &                     &                     & \\ 
\hline
$T_{\text {int}}$ (K)        &A: 300                  & 182$^{+87}_{-88}$  & 185$^{+100}_{-90}$    &176$^{+94}_{-87}$     &294$^{+19}_{-40}$      &       -              &       -             &      -              &-\\
                          &B: 200                  &      -               &            -         &   -                  &         -             &206$^{+110}_{-104}$    &132$^{+73}_{-57}$  &205$^{+118}_{-108}$  &212$^{+21}_{-20}$  \\
\hline
$T_{\text{equi}}$ (K)         &A: 1200                 &  1174$^{+69}_{-58}$  &  1558$^{+28}_{-38}$  &  1173$^{+72}_{-52}$  &  1283$^{+49}_{-185}$   &            -         &       -             &         -           &-\\
                          &B: 1000                 &             -        &        -             &       -              &          -            &  985$^{+15}_{-14}$    &  800$^{+1}_{-1}$   &  1074$^{+14}_{-9}$  &  920$^{+61}_{-60}$  \\
\hline
$\log_{10}(K_{zz}/\text{cm}^2 \text{ s}^{-1})$ &A1: 0, A2: 9&  -           &  -                   &2.75$^{+1.98}_{-1.80}$&10.51$^{+0.72}_{-3.34}$&        -             &   -                 &  -  & - \\
                          &B1: 0, B2: 9             &           -          &          -           &          -           &          -            &           -          &          -            & 0.25$^{+0.45}_{-0.18}$  &8.52$^{+2.43}_{-2.96}$\\
\hline
$\log_{10}(g/\text{cm s}^{-2})$   & 3              &2.99$^{+0.03}_{-0.03}$&3.11$^{+0.02}_{-0.02}$&2.99$^{+0.03}_{-0.02}$&3.03$^{+0.02}_{-0.08}$ &2.99$^{+0.02}_{-0.02}$&2.78$^{+0.01}_{-0.02}$&3.08$^{+0.02}_{-0.02}$&2.96$^{+0.04}_{-0.04}$\\
\hline
[M/H] (dex)               &A: 1                   &0.99$^{+0.07}_{-0.07}$&0.86$^{+0.05}_{-0.04}$&0.98$^{+0.07}_{-0.07}$&0.96$^{+0.08}_{-0.07}$  &                      &                      &                      &                      \\
                          &B: 0                   &                      &                      &                      &                        &0.03$^{+0.04}_{-0.05}$&0.36$^{+0.03}_{-0.04}$&0.05$^{+0.04}_{-0.04}$&-0.00$^{+0.06}_{-0.06}$\\
\hline
$\chi^{2}$               &        -               &  1.10                &  1.29                &       1.11           &       1.15             &     01.09               &     1.97           &    1.15                     & 1.03\\ 
\hline
\end{tabular}
\end{center}
\end{table}

\subsection{Transit Retrieval Results}
In this section, we discuss the result of the retrieval analysis of the four test cases, namely A1, A2, B1, and B2. We have run two retrieval models for each synthetic JWST observation: the quench level retrieval model (QLR) and the chemically consistent retrieval model (CCR). In total, we have run eight transmission retrieval models (CCR:[CCR-A1, CCR-A2, CCR-B1, CCR-B2] and QLR:[QLR-A1, QLR-A2, QLR-B1, QLR-B2]) to understand the potential capability of the QLR. We used pre-computed thermochemical abundance data for the CCR and interpolated 
the data for the input thermal profile. For all our retrieval models, we initialized 1000 initial samples in which the prior parameters are distributed as given in Table \ref{Table5_5}. Then, we ran 20,000 models around the initial samples' best-fit model. The outputs of these model runs are used to constrain the atmospheric parameters statistically. Tables \ref{Table5_3} shows the retrieved parameters for the cases A and B and the true values (the initial parameters used to calculate the atmospheric composition as given in Table \ref{Table5.1}).

\begin{figure}[t!]
\centering
\includegraphics[trim={2cm 2cm 4.5cm 2cm},clip, width=0.49\linewidth]{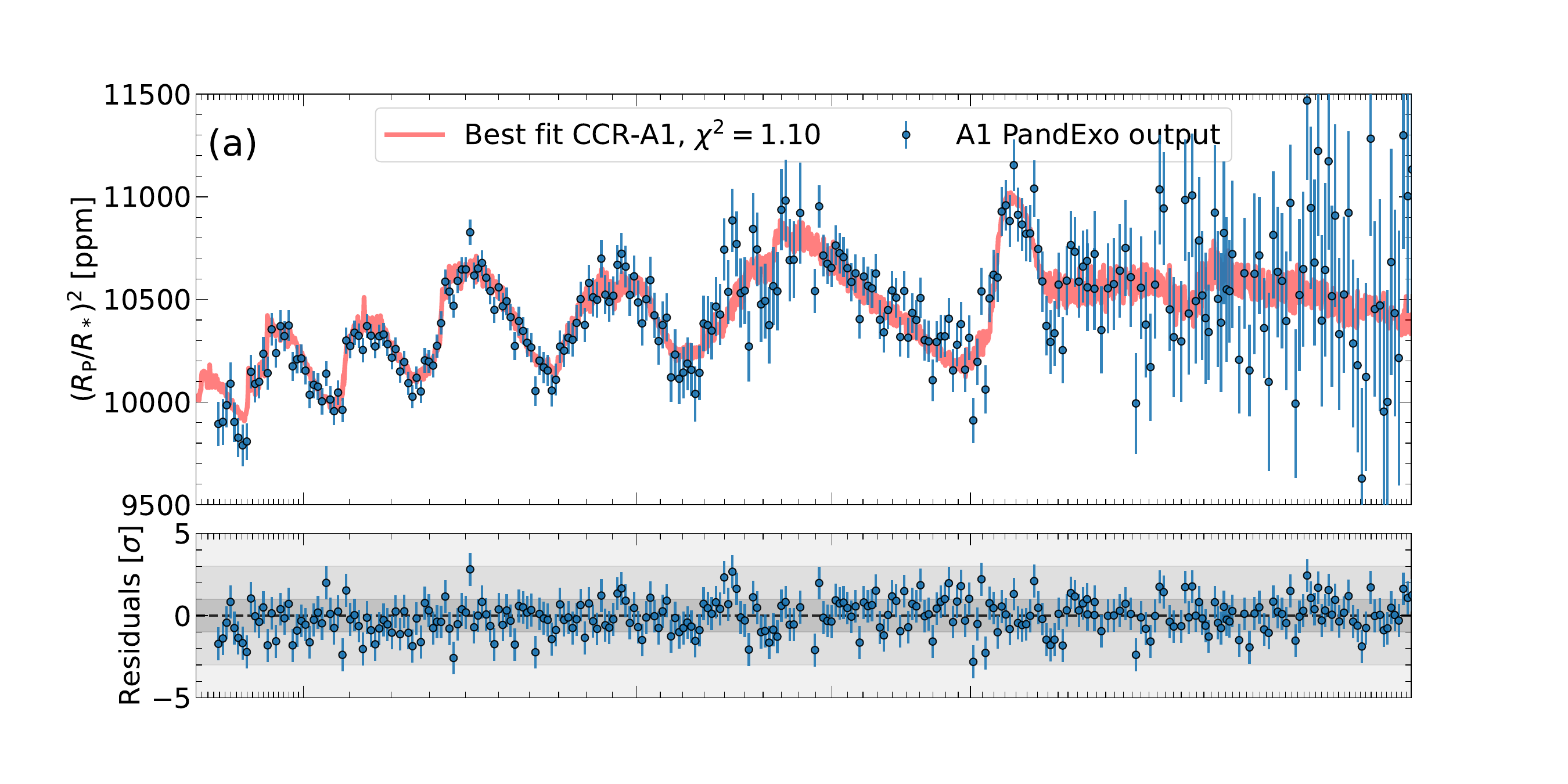}
\includegraphics[trim={4cm 2cm 2.5cm 2cm},clip, width=0.49\linewidth]{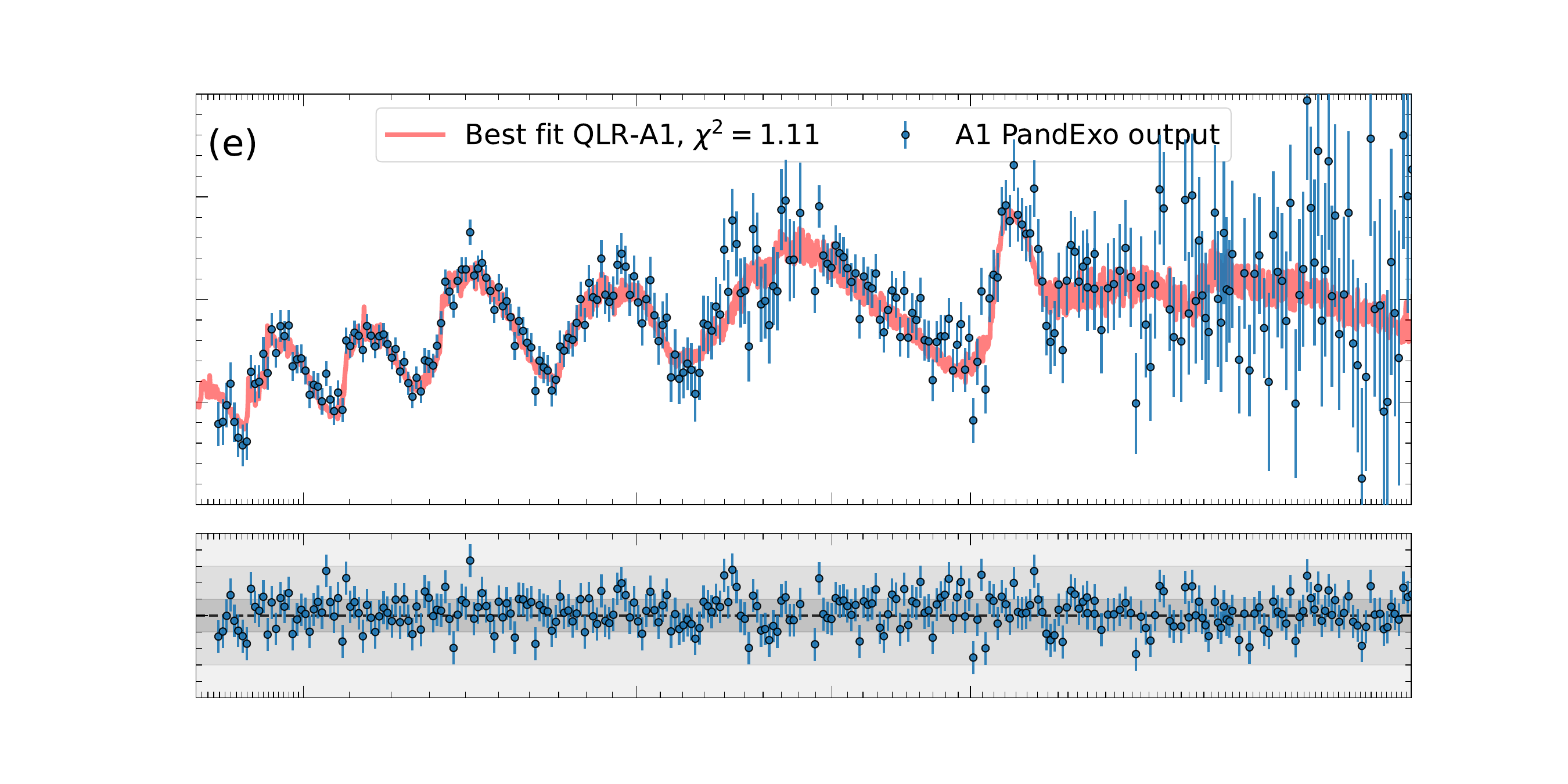}
\includegraphics[trim={2cm 2cm 4.5cm 2cm},clip, width=0.49\linewidth]{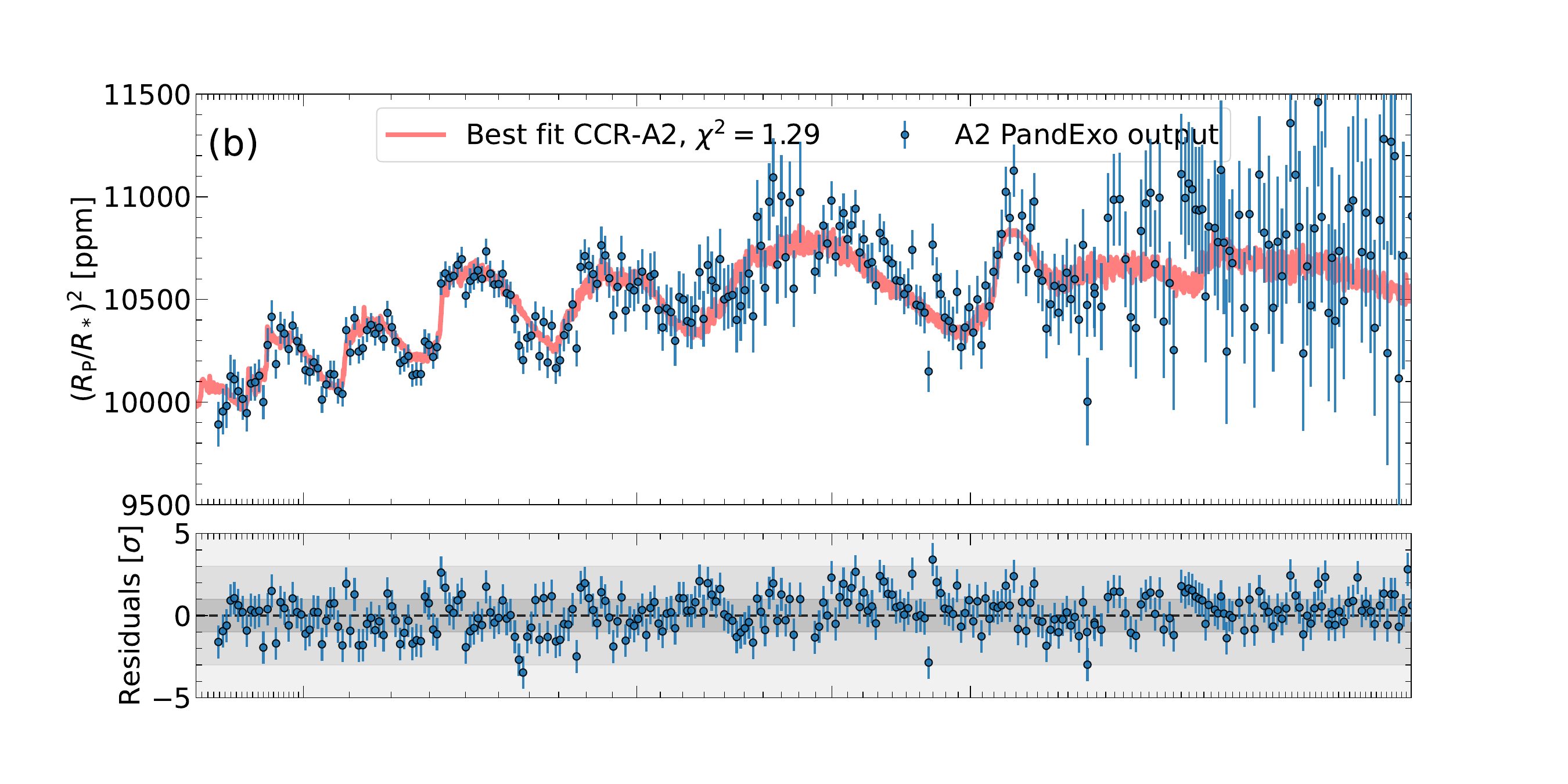}
\includegraphics[trim={4cm 2cm 2.5cm 2cm},clip, width=0.49\linewidth]{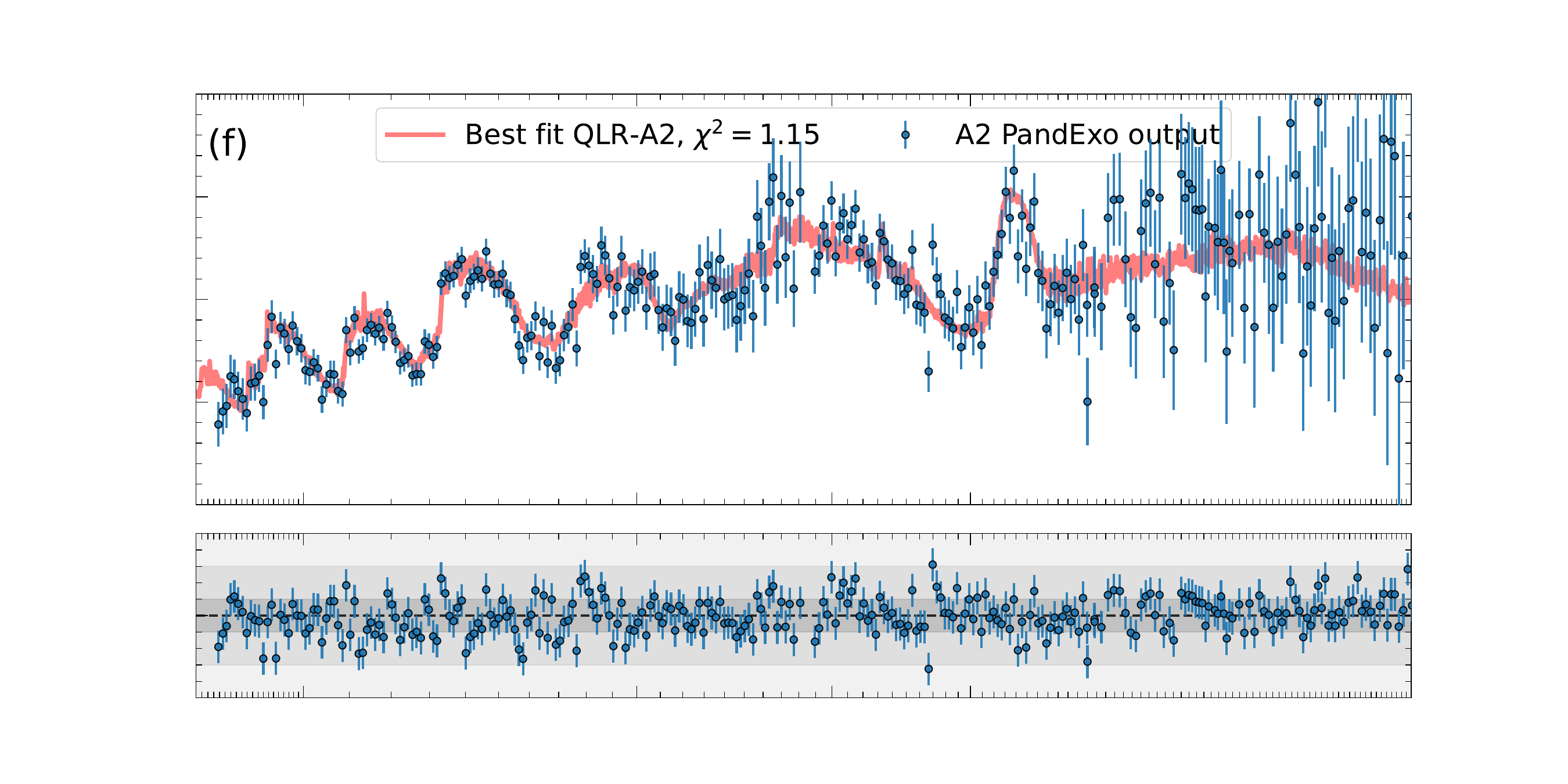}
\includegraphics[trim={2cm 2cm 4.5cm 2cm},clip, width=0.49\linewidth]{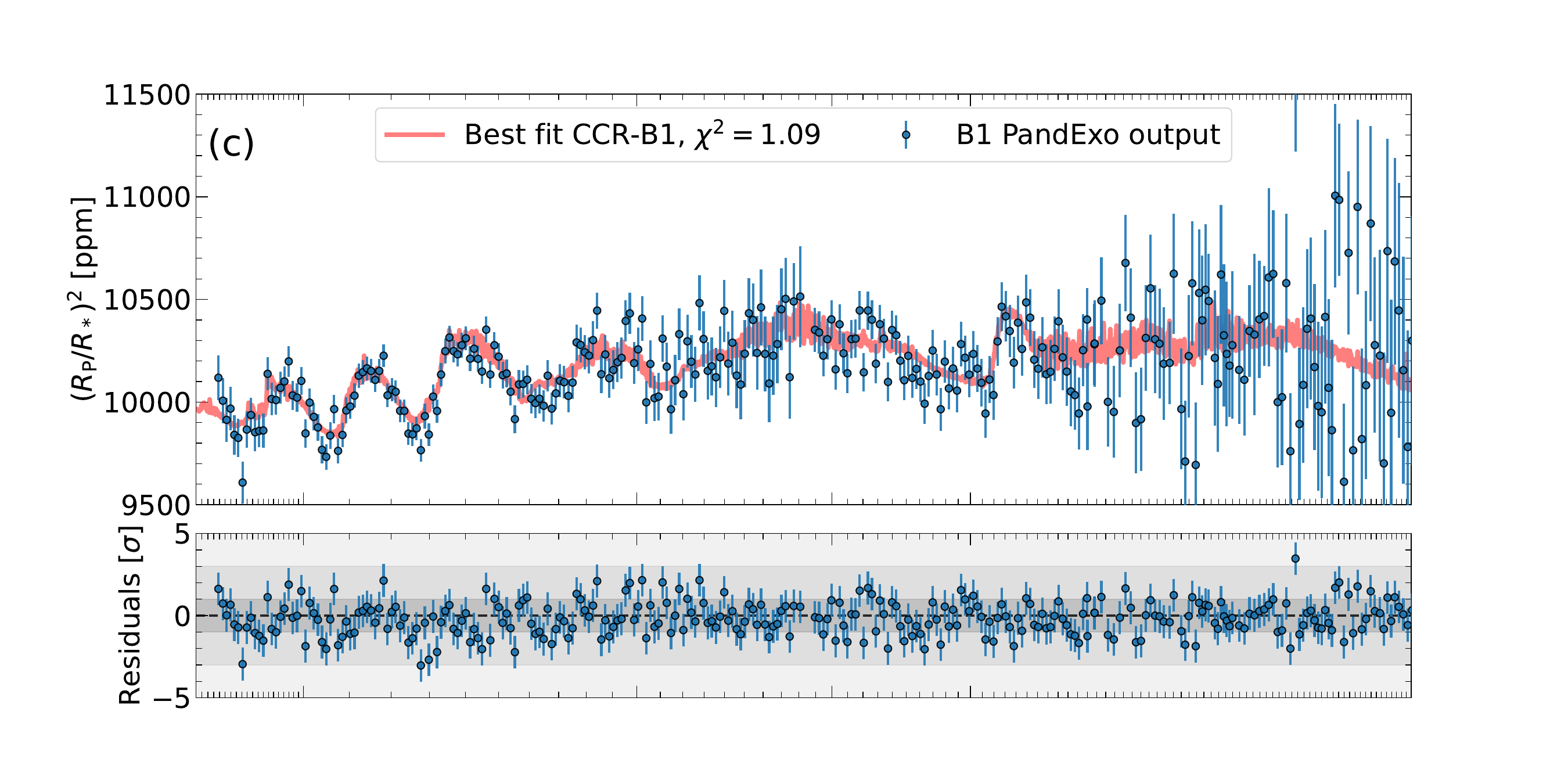}
\includegraphics[trim={4cm 2cm 2.5cm 2cm},clip, width=0.49\linewidth]{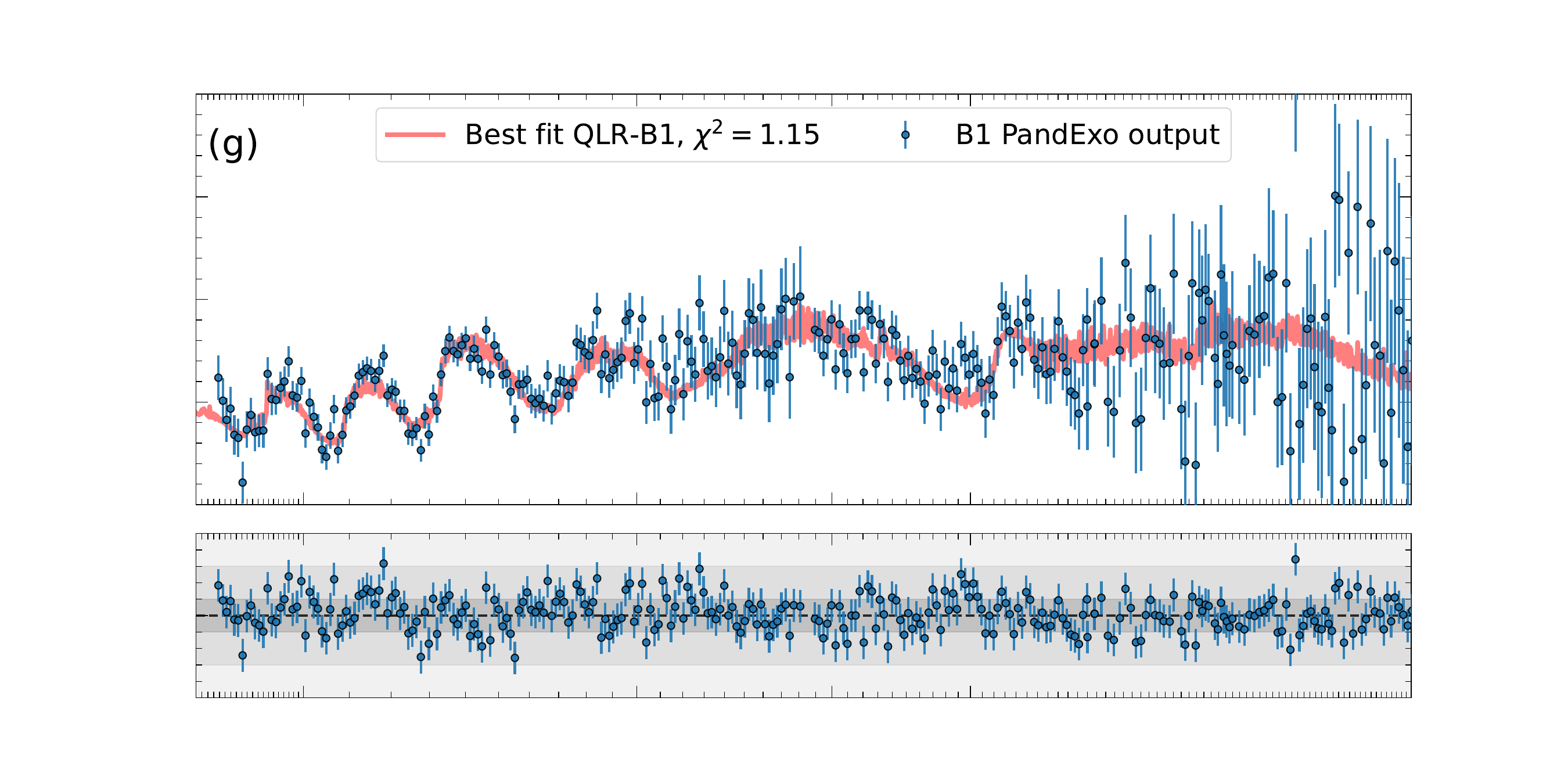}
\includegraphics[trim={2cm 0cm 4.5cm 2cm},clip, width=0.49\linewidth]{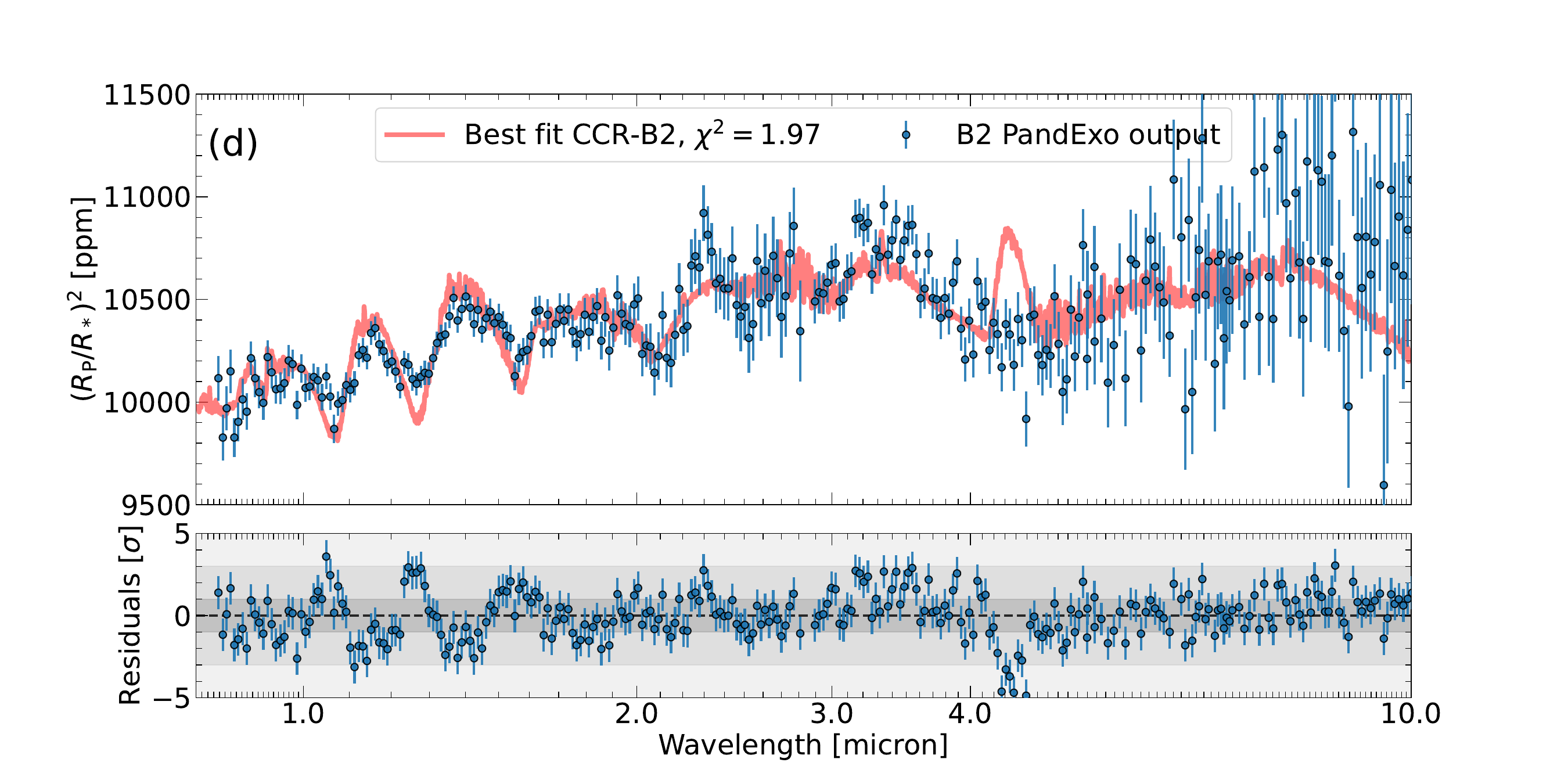}
\includegraphics[trim={4cm 0cm 2.5cm 2cm},clip, width=0.49\linewidth]{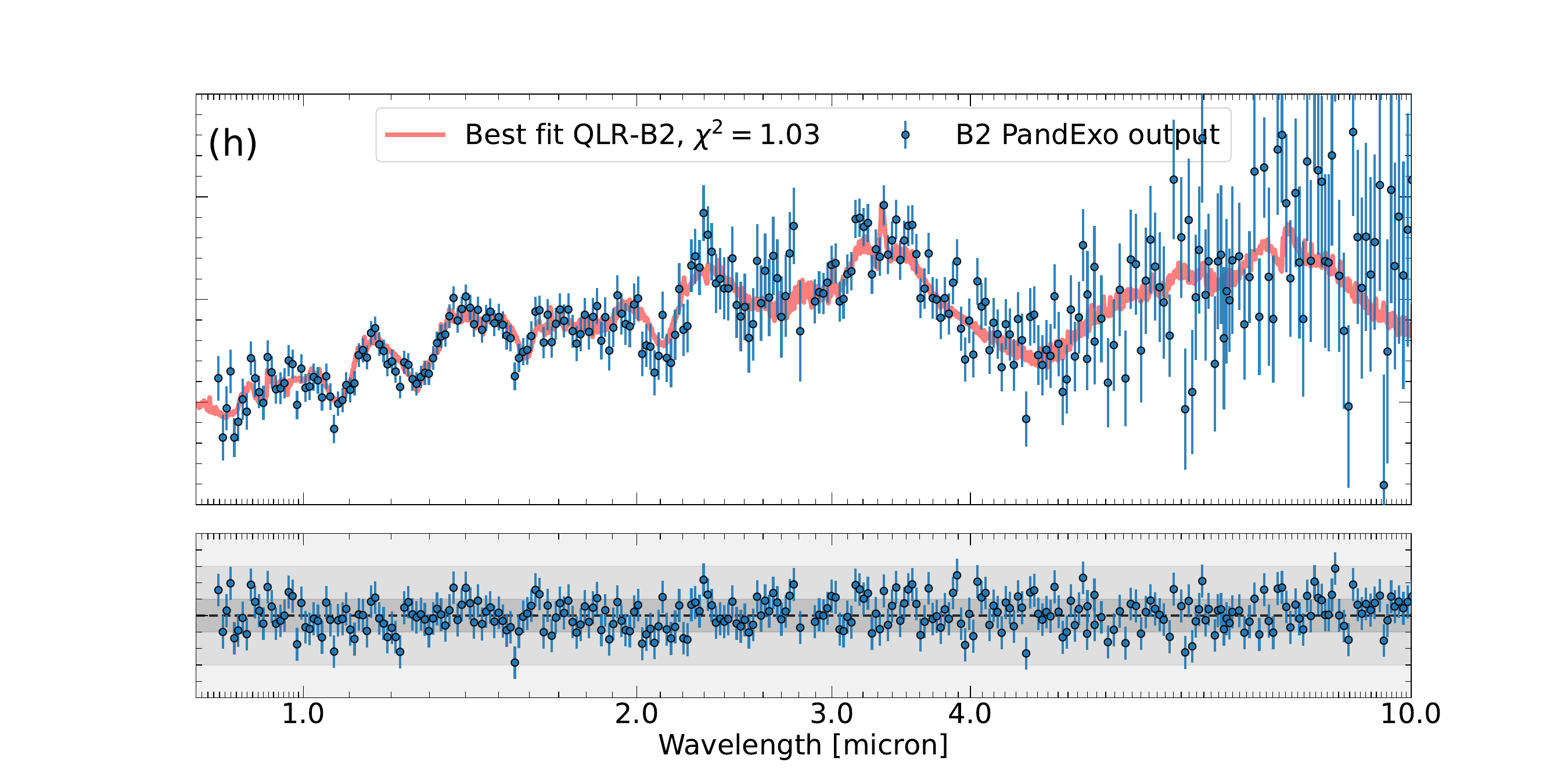}
\caption{The best-fit model transit spectrum generated from the CCR (a-d) and QLR (e-h) models for the cases A1 (first row), A2 (second row), B1 (third row), and B2 (fourth row) for the synthetic JWST observation of the respective model run.}
\label{Fig:QLR_CCR_A1_A2_best_fit}
\end{figure}

\begin{figure}[t!]
\centering
\includegraphics[width=0.45\linewidth]{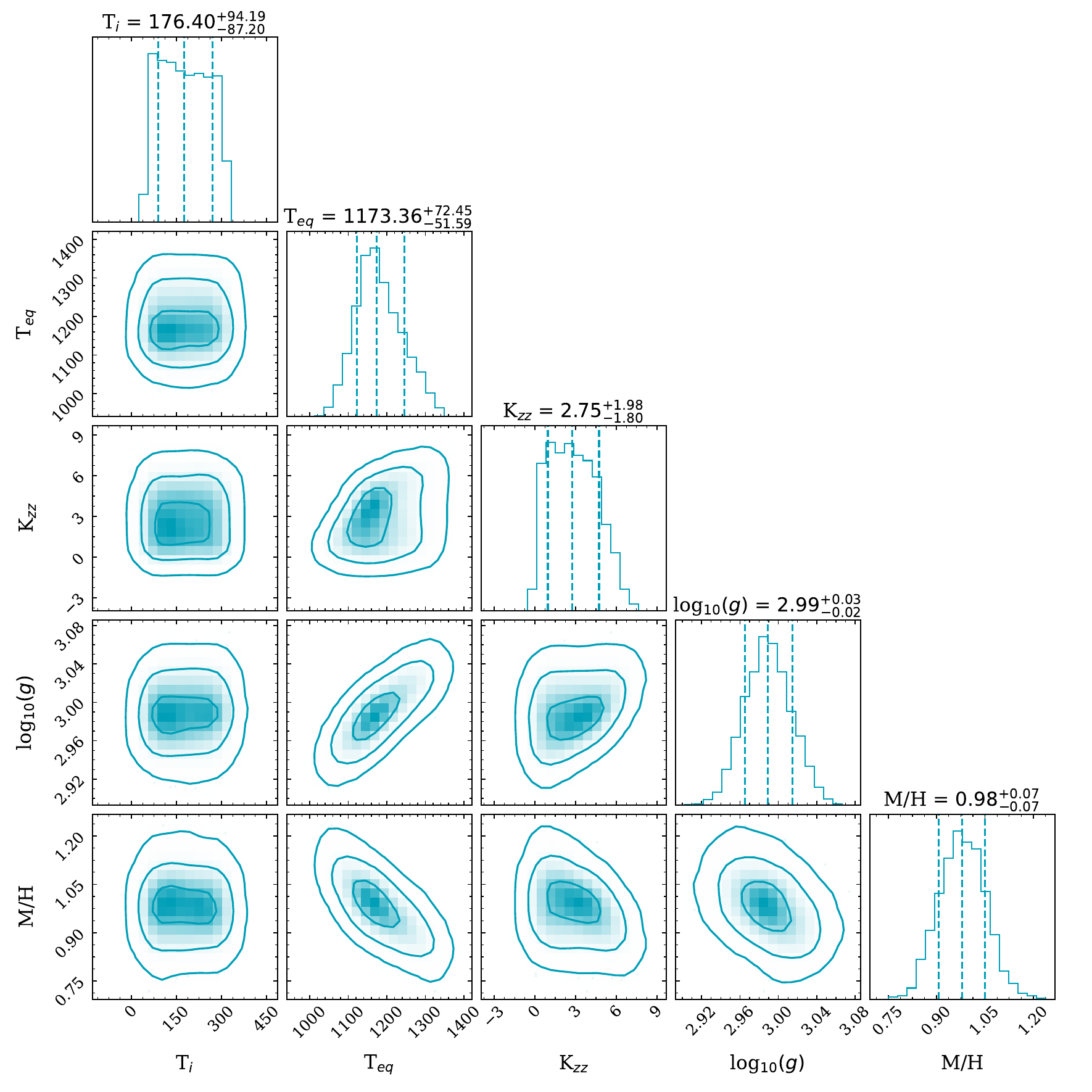}
\includegraphics[width=0.45\linewidth]{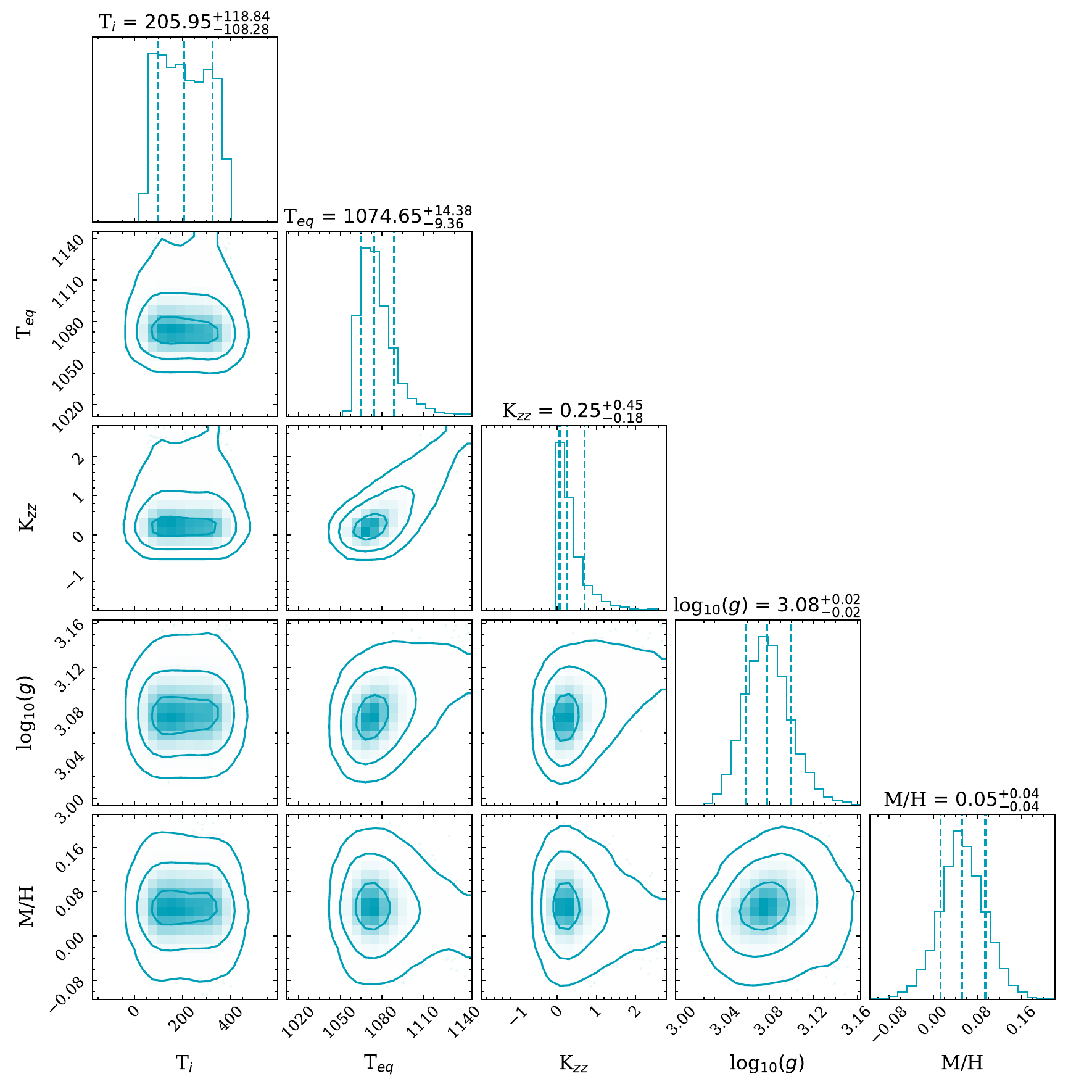}
\includegraphics[width=0.45\linewidth]{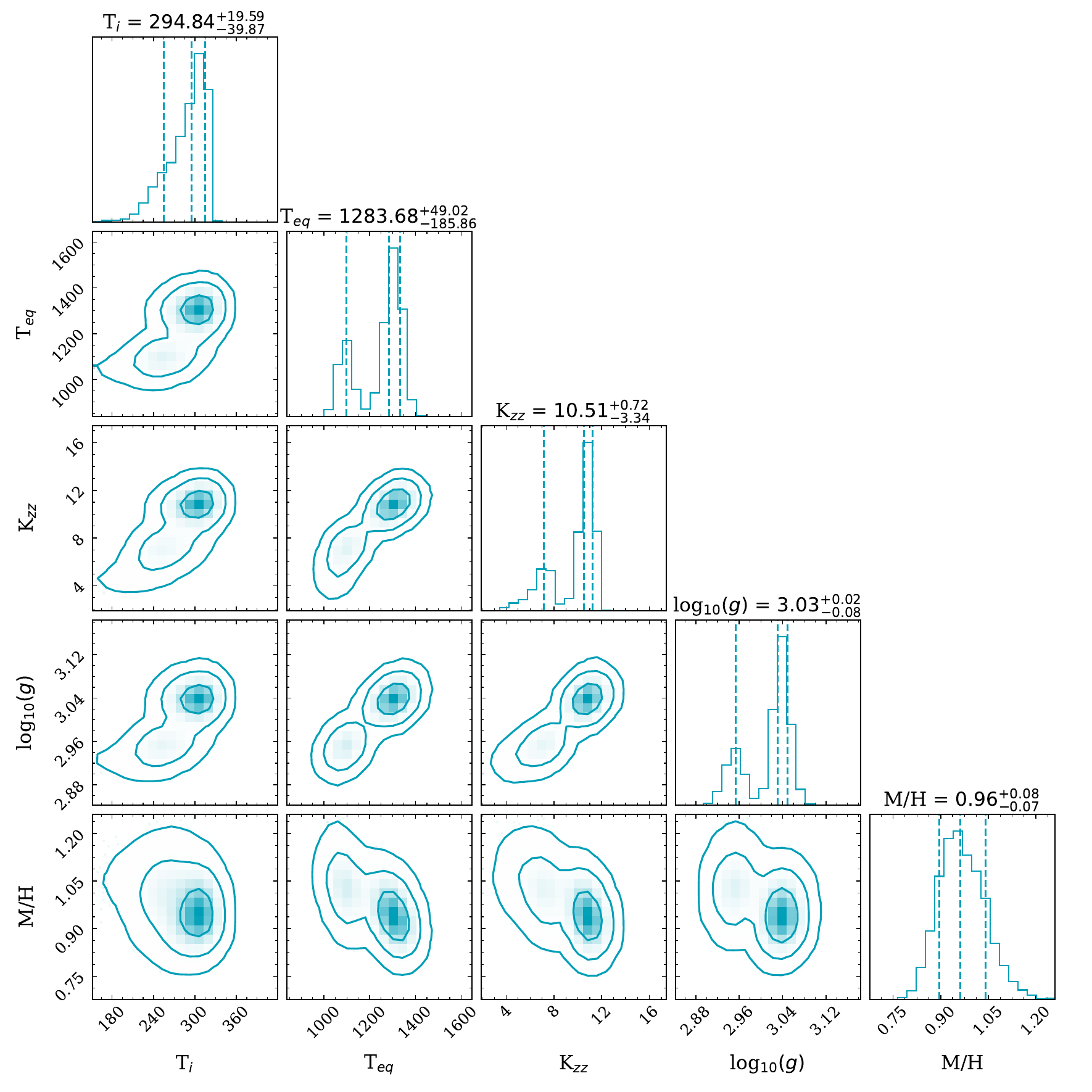}
\includegraphics[width=0.45\linewidth]{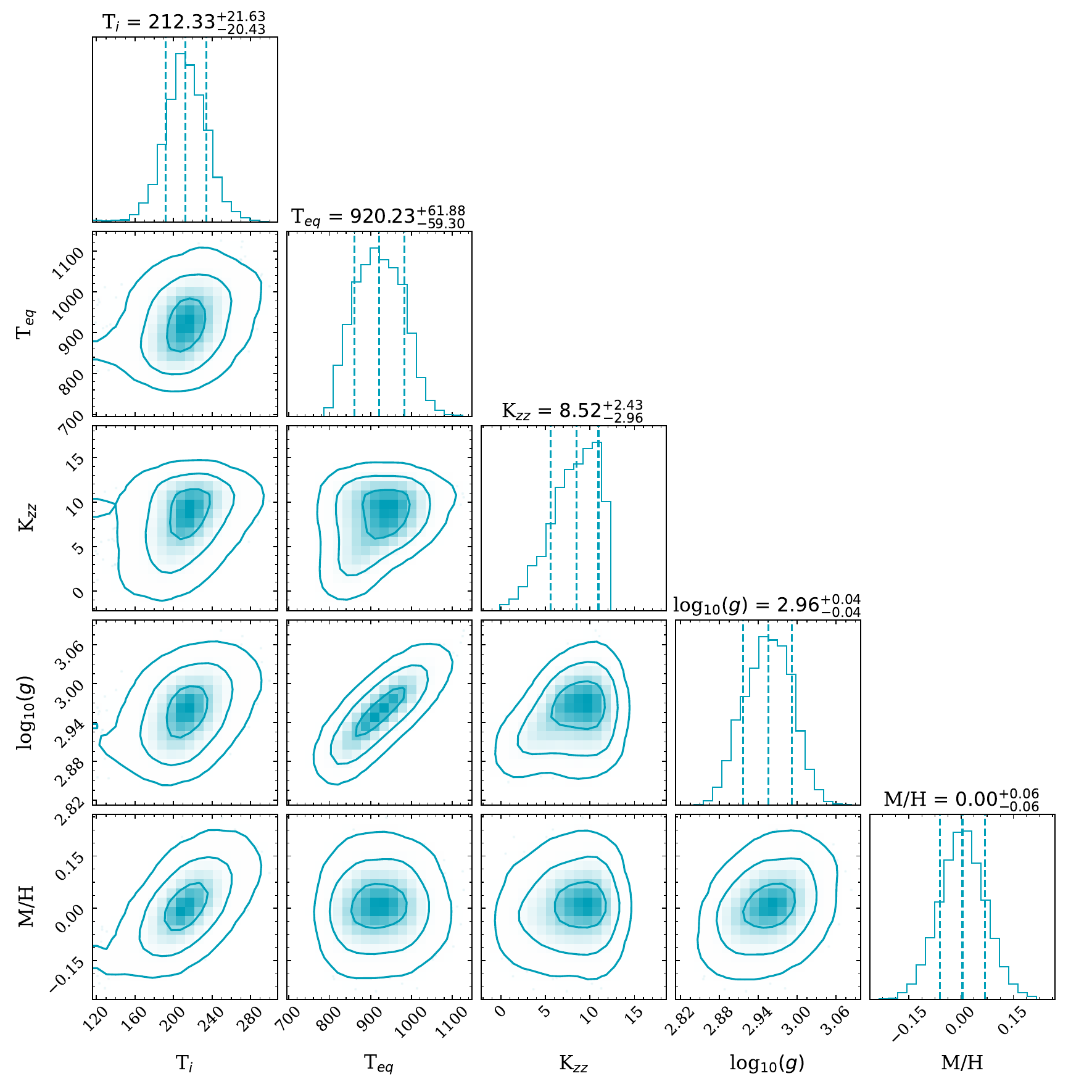}
\caption{The figure shows the corner plot (posterior distribution) of the parameters retrieved for model run from QLR. A1: top left, A2: bottom left, B1: top right, B2: bottom right. T$_i$ is the internal temperature and T$_{eq}$ is the equilibrium temperature.}
\label{Fig:QLR_cornor}
\end{figure}

We have summarized our quench level retrieval model (QLR) in Figures \ref{Fig:QLR_CCR_A1_A2_best_fit}(e) - (h) for the best fit modeled spectra of QLR-A1, QLR-A2, QLR-B1, and QLR-B2, and Figure \ref{Fig:QLR_cornor} for the posterior distributions. These figures demonstrate an excellent fit, with residuals scattered around zero and mostly within 1$\sigma$. The $\chi^2$ values for the equilibrium models QLR-A1 and QLR-B1 are 1.11 and 1.15, respectively, while for the disequilibrium models QLR-A2 and QLR-B2, they are 1.15 and 1.03, respectively. 
The retrieved internal temperature for QLR-A1 and QLR-B1 are $176^{+94}_{-87}$ K and $205^{+118}_{-108}$ K, respectively. Since the transit spectra of these cases are independent of internal temperature, moderate changes in a planet’s internal temperature do not significantly impact the thermal profile in the infrared photosphere, leaving the thermochemical equilibrium composition in this region unaffected. The retrieved equilibrium temperatures for QLR-A1 and QLR-B1 are $1173^{+72}_{-52}$ K and $1074^{+14}_{-9}$ K, respectively, and are better constrained than the internal temperatures, as the equilibrium temperature directly influences the transit spectrum in the thermochemical equilibrium cases (A1 and B1).

In cases QLR-A2 and QLR-B2, the thermal profile at higher pressure levels (higher pressure thermal profile largely affected by the internal temperature) influences both the quench pressure level and the quenched abundances of species. The internal temperature is more tightly constrained in QLR-A2 and QLR-B2 compared to QLR-A1 and QLR-B1. The retrieved values of $T_{\text{int}}$ for QLR-A2 and QLR-B2 are $294^{+19}_{-40}$ K and $212^{+21}_{-20}$ K, respectively. The retrieved equilibrium temperatures ($T_{\text{equi}}$) for QLR-A2 and QLR-B2 are $1283^{+40}_{-185}$ K and $920^{+61}_{-60}$ K, respectively. The retrieved values of the eddy diffusion coefficient $\log_{10}(K_{zz})$ for the equilibrium cases QLR-A1 and QLR-B1 are $2.8^{+2}_{-1.8}$ and $0.3^{+0.5}_{-0.2}$, respectively, indicating that thermochemical equilibrium is dominant. In the disequilibrium cases QLR-A2 and QLR-B2, the retrieved values of $\log_{10}(K_{zz})$ are $10.5^{+0.7}_{-3.3}$ and $8.5^{+2.4}_{-3.0}$, respectively. Across all test cases of QLR, surface gravity and atmospheric metallicity are constrained within $1\%$ of their true values.

\begin{figure}[t!]
\centering
\includegraphics[width=0.45\linewidth]{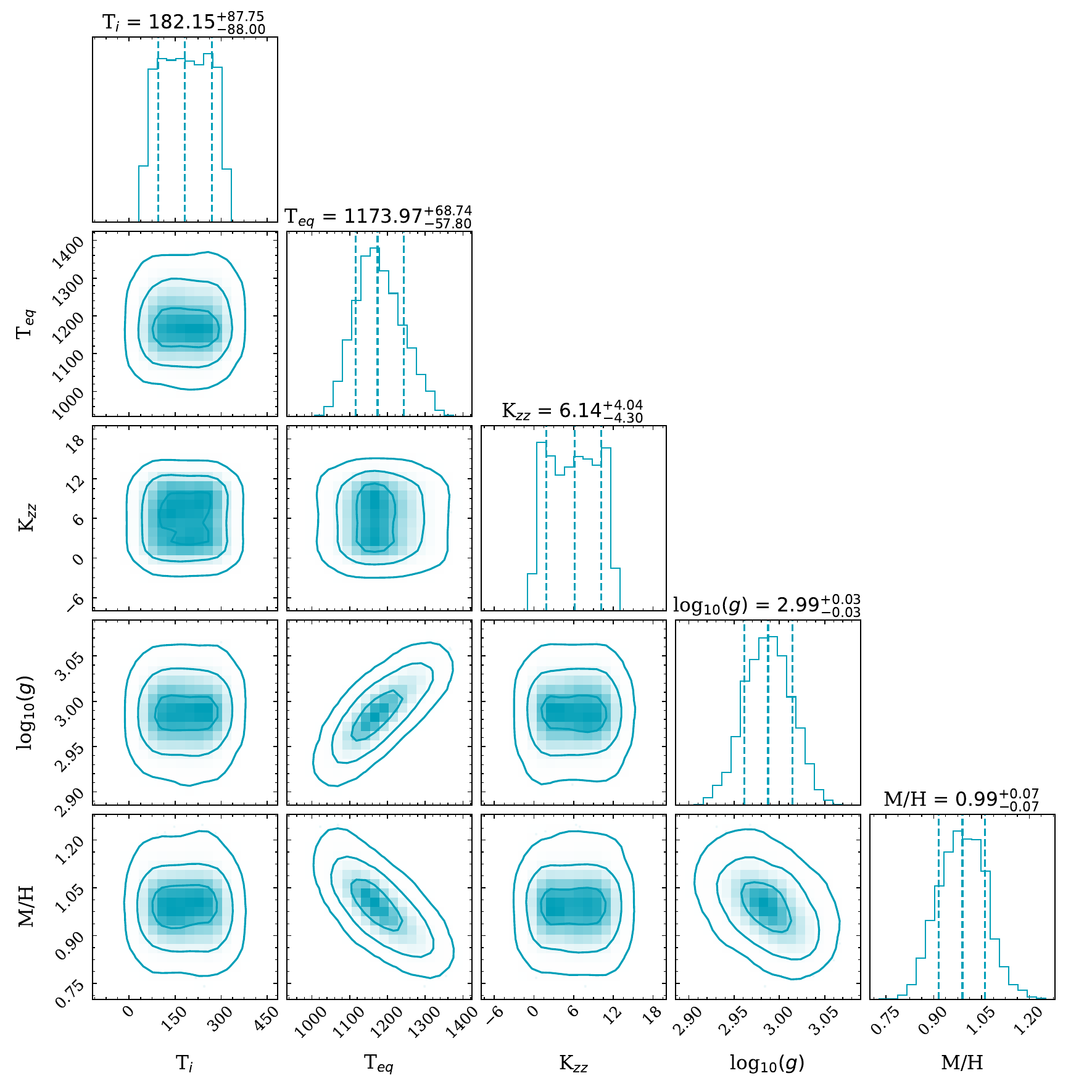}
\includegraphics[width=0.45\linewidth]{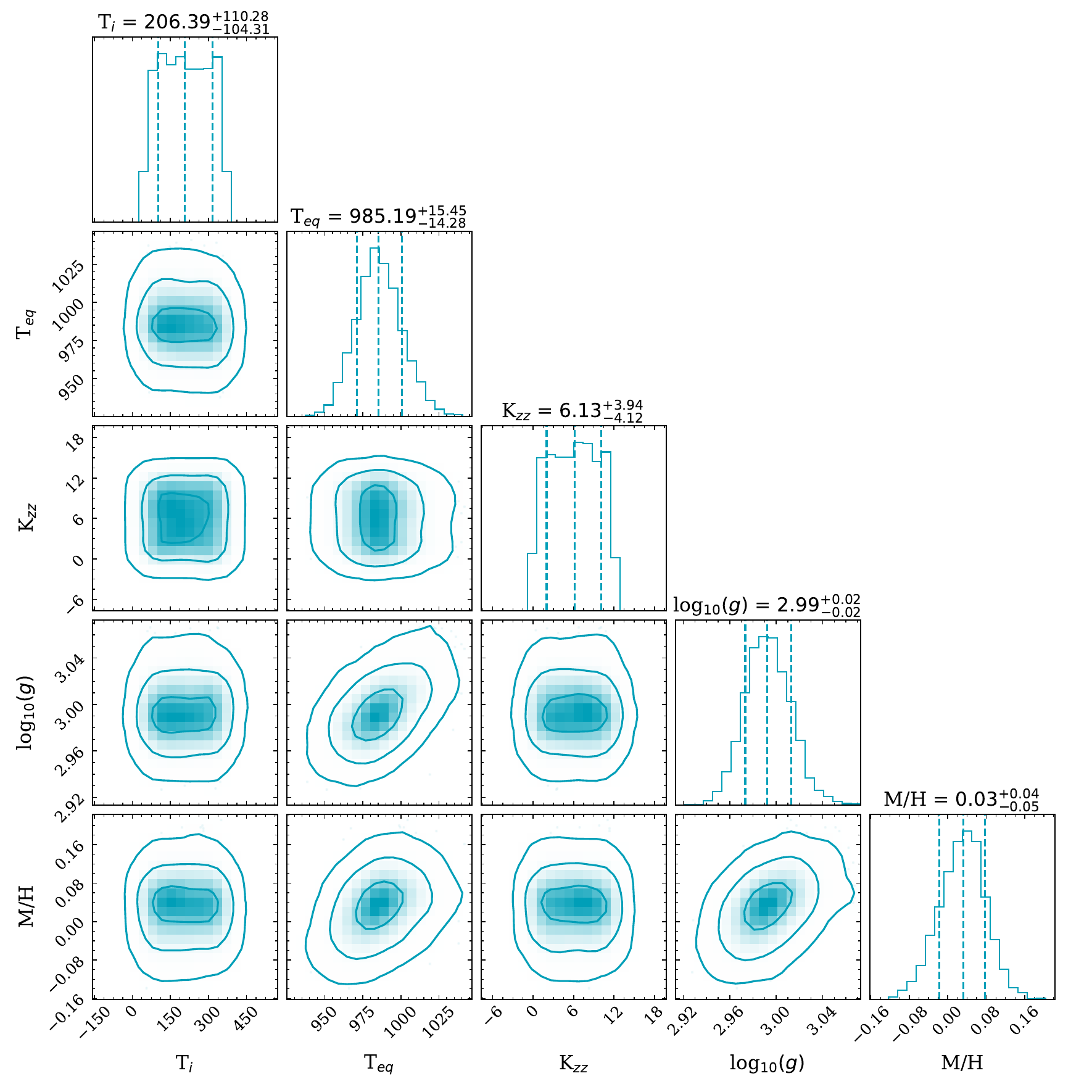}
\includegraphics[width=0.45\linewidth]{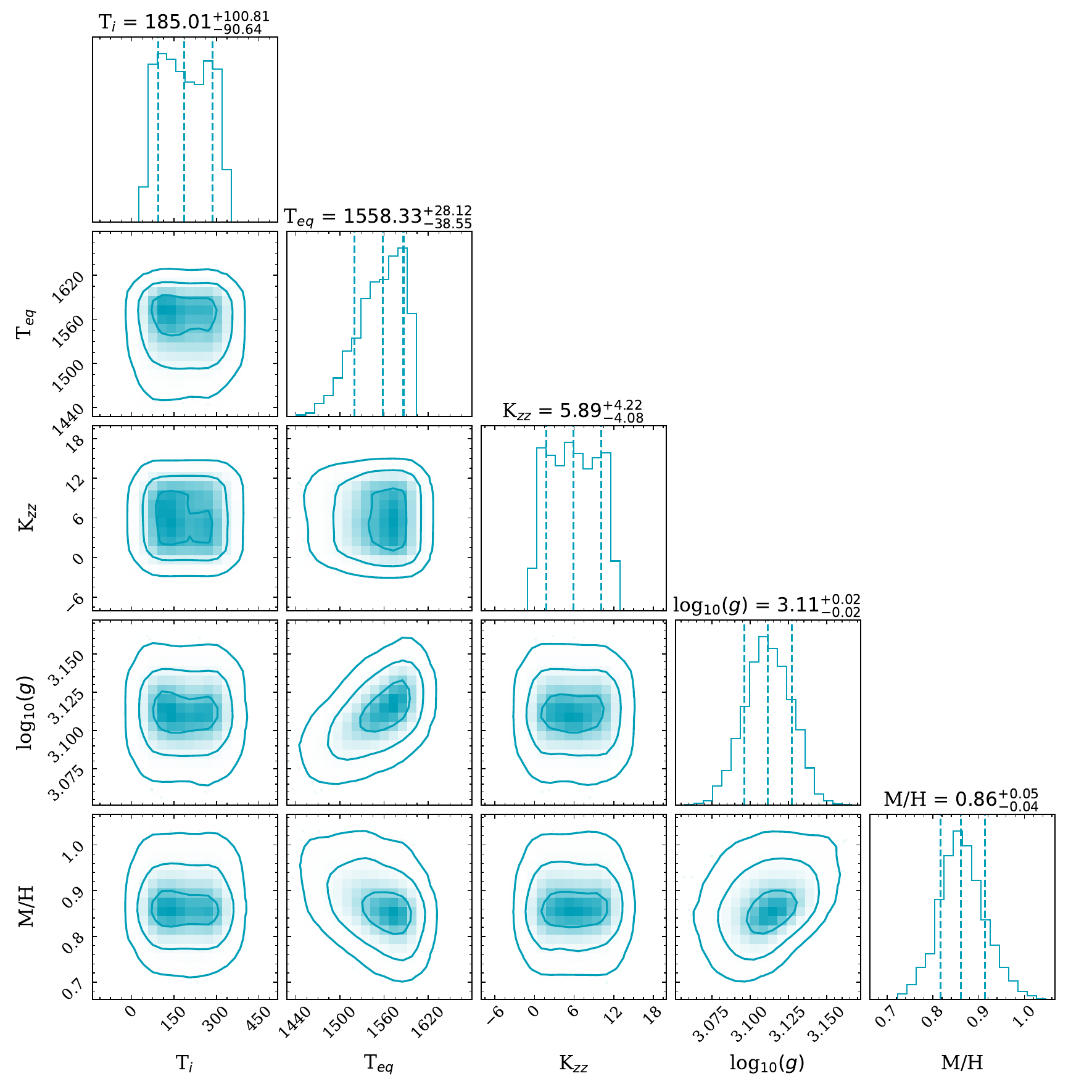}
\includegraphics[width=0.45\linewidth]{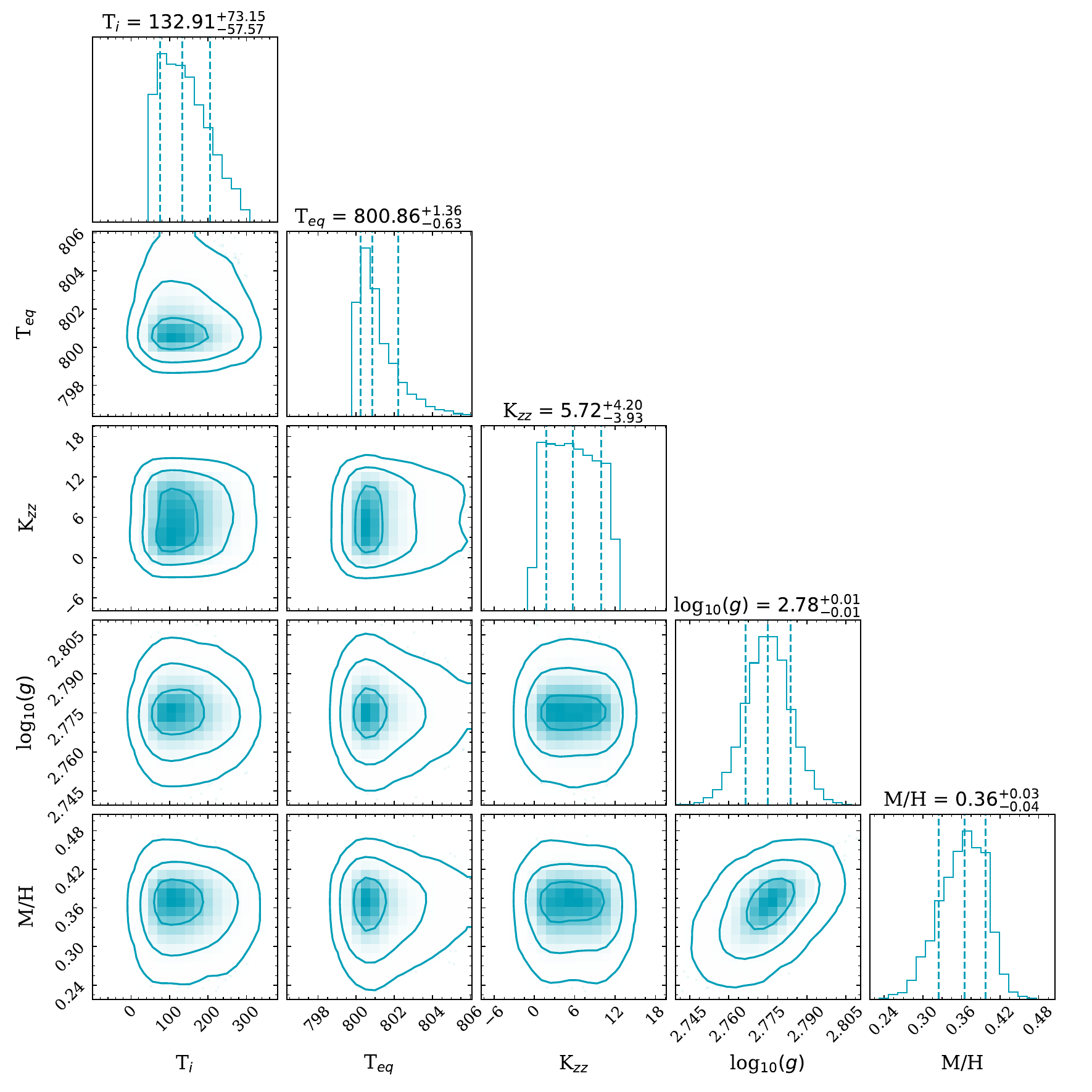}
\caption{The figure shows the corner plot (posterior distribution) of the parameters retrieved for model run from CCR. A1: top left, A2: bottom left, B1: top right, B2: bottom right. T$_i$ is the internal temperature and T$_{eq}$ is the equilibrium temperature.}
\label{Fig:CCR_cornor}
\end{figure}

 The results of the chemically consistent retrieval model (CCR) are summarized in Figures  \ref{Fig:QLR_CCR_A1_A2_best_fit}(a) - (d) for the best fit modeled spectra of CCR-A1, CCR-A2, CCR-B1, and CCR-B2 and Figure \ref{Fig:CCR_cornor} for the posterior distributions. The fit is good for CCR-A1 and CCR-B1; the residuals are scattered around zero and within 1$\sigma$ error. However, in the case of CCR-B2, the residuals do not scatter evenly around zero, and the residuals are outside 1$\sigma$. The $\chi^2$ for the equilibrium models CCR-A1 and CCR-B1 are 1.10 and 1.09, respectively, and for the disequilibrium models CCR-A2 and CCR-B2, the $\chi^2$ values are 1.29 and 1.97, respectively. Thus, it is evident that the QLR models are more suitable than the CCR models, where vertical mixing dominates the atmosphere.
The retrieved parameter range of $T_{\text{int}}$ and $T_{\text{equi}}$ for the case CCR-A1 and CCR-B1 are similar to the QLR method and close to their actual values. The maximum deviation of the retrieved $T_{\text{equi}}$ from the actual value is for the CCR-A2 and CCR-B2 around 258 K and 200 K, respectively. The retrieved values of [M/H] (atmospheric metallicity) for the cases CCR-A1 and CCR-B1 are close to their true values and similar to the value given in the QLR method. Whereas, for the cases CCR-A2 and CCR-B2, the deviation of the retrieved atmospheric metallicity with the true values are 0.14 and 0.36, resepectively.

\clearpage

\subsection{Summary}
We use a custom-built disequilibrium model function in the petitRADTRANS python module to make a quench level retrieval method (QLR). We have compared QLR with the chemically consistent retrieval model (CCR). For this, we use two test cases, A and B, in which B shows a strong disequilibrium effect on its atmospheric abundance compared to A. The following is a comparison of the results of QLR and CCR.

\begin{itemize}
	\item  The retrieved $\log_{10}(K_{zz})$ values for the equilibrium test cases (QLR-A1 and QLR-B1) are small, indicating a lack of disequilibrium chemistry.
	\item  The best-fit retrieved $\log_{10}(K_{zz})$ values for the disequilibrium test cases (QLR-A2 and QLR-B2) are within 15\% of the true values.
	\item  For most cases, the best-fit atmospheric metallicity ([M/H]) closely matches the true values, with the exception of the CCR-A2 and CCR-B2 model outputs.
	\item  The $\chi^2$ value is significantly large for the CCR-B2 model run.
	\item  For disequilibrium test cases, QLR-A2 and QLR-B2 provide much better constraints on atmospheric parameters compared to CCR-A2 and CCR-B2. In equilibrium test cases, QLR-A1 and QLR-B1 perform similar to CCR-A1 and CCR-B1.
\end{itemize}

\end{document}